\documentclass[times,doublespace]{simauth}

\usepackage{moreverb}

\usepackage[colorlinks,bookmarksopen,bookmarksnumbered,citecolor=red,urlcolor=red]{hyperref}

\newcommand\BibTeX{{\rmfamily B\kern-.05em \textsc{i\kern-.025em b}\kern-.08em
T\kern-.1667em\lower.7ex\hbox{E}\kern-.125emX}}

\def\volumeyear{2018}

\usepackage{lscape,rotating}
\usepackage{float,epsfig}
\usepackage{graphicx, caption, subcaption}
\usepackage{setspace}
 \usepackage{amsmath,bm} 
\usepackage{amsfonts}
\usepackage{amssymb}

\newcommand{\xv}{\bm{x}}
\newcommand{\yv}{\bm{y}}
\newcommand{\zv}{\bm{z}}

\newcommand{\wc}{\text{\small$\mathcal{W}$}}
\newcommand{\Ivv}{\bm{\text{I}}}
\newcommand{\Uvv}{\bm{\text{U}}}

\newcommand{\Deltav}{\bm{\Delta}}

\newcommand{\piv}{\bm{\pi}}

\newcommand{\Thetav}{\bm{\Theta}}

\newcommand{\betav}{\bm{\beta}}

\newcommand{\zerov}{\bm{0}}

\makeatletter
\long\def\@makecaption#1#2{%
  \vskip\abovecaptionskip
  \sbox\@tempboxa{#1: #2}%
  \ifdim \wd\@tempboxa >\hsize
    #1: #2\par
  \else
    \global \@minipagefalse
    \hb@xt@\hsize{\box\@tempboxa\hfil}%
  \fi
  \vskip\belowcaptionskip}
\makeatother

\begin{document}

\runninghead{Y. Tang}

\title{A monotone data augmentation algorithm for multivariate nonnormal data: with applications to controlled imputations for longitudinal  trials}

\author{Yongqiang Tang}

\address{Shire, 300 Shire Way, Lexington, MA 02421, USA}

\corraddr{E-mail: yongqiang\_tang@yahoo.com}

\begin{abstract}
An efficient monotone data augmentation (MDA) algorithm is proposed  for missing data imputation for incomplete multivariate nonnormal data that may contain variables of 
different types, and are modeled by a sequence of regression models including the linear, binary logistic, multinomial logistic, 
proportional odds, Poisson, negative binomial, skew-normal, skew-t  regressions or a mixture of these  models.
The MDA algorithm is applied to the sensitivity analyses of longitudinal  trials with nonignorable dropout using the controlled pattern imputations that assume 
the treatment effect reduces or disappears after subjects in the experimental arm discontinue the treatment.
We also  describe a heuristic approach to implement  the controlled imputation, in which  the  fully conditional specification method is used to impute the intermediate missing data to
create a monotone missing pattern, and the missing data after dropout are then  imputed
according to the assumed nonignorable mechanisms.
The proposed methods are illustrated by simulation and real data analyses.
 \end{abstract}

\maketitle

\section{Introduction} 
Multiple imputation (MI) provides a popular and convenient way to analyze complex data with missing values \cite{rubin:1996}. A MI procedure consists of three steps: 1)
The missing values are imputed $m$ times from their posterior predictive distribution given the observed data on basis of an appropriate statistical model; 2) Each imputed 
dataset is analyzed by a standard statistical method; 3) The results from $m$ datasets are combined for inference by using Rubin's rule \cite{rubin:1987}.
An attractive feature of MI is that the imputation and analysis models can be different. For example, in  clinical trials, the surrogate endpoints and 
auxiliary variables are often highly correlated with the primary efficacy endpoint and the dropout process, and may be employed  to improve the imputation of the primary 
efficacy outcomes \cite{rubin:1996, little:1996, faucett:2002, collins:2001, schafer:2003}, but it is difficult to incorporate such information  in the likelihood-based inference \cite{collins:2001,schafer:2003}.

The MI methodology  is well established for  multivariate normal outcomes with an arbitrary missing pattern.  An efficient Markov chain Monte Carlo (MCMC) algorithm was developed by  Schafer \cite{schafer:1997} 
by using the monotone data augmentation (MDA) technique. The MDA algorithm iterates between an imputation I-step, in which the intermittent missing data are imputed given the current draw of the model parameters, 
and a posterior P-step, in which the model parameters are updated  given the current imputed monotone data. It tends to converge faster with smaller autocorrelation between posterior samples than a full data augmentation 
algorithm that imputes both the intermittent missing data and missing data after dropout during the I-step \cite{schafer:1997,2016:tang}.  Schafer's algorithm
was recently improved by Tang  \cite{2015a:tang,2016:tang}. 
 Tang's approach  allows the use of a
more general prior distribution \cite{2015a:tang,2016:tang},   imputes the intermittent missing outcomes in a more computationally efficient way  \cite{2016:tang},
and enables more flexible modeling of the mean and covariance matrix \cite{2016:tang}. 

For multivariate nonnormal data with a monotone missing pattern, imputation can be performed by  the sequential regression method  \cite{raghunathan:2001}.
The multivariate nonnormal data with an arbitrary missing pattern are generally imputed by  the MCMC method for multivariate normal outcomes \cite{Bernaards:2007, lee:2010b,donneau:2015a} or by the 
 fully conditional specification (FCS) method \cite{raghunathan:2001, buuren:2006,buuren:2007, lee:2010b, white:2011} due to the lack of a natural multivariate distribution for these data. The former approach ignores
  the non-normality in the data. The FCS, also known as ``chained equations'', is an analogy to the 
traditional Gibbs sampling scheme \cite{casella:1992}, and imputes the data on a variable-by-variable basis by specifying a conditional model for each variable with all other variables as predictors. 
A theoretical weakness of FCS is that there does not in general  exist a joint distribution that is consistent with these  conditional distributions  \cite{buuren:2007,liu:2014,chen:2015}, 
and its performance is evaluated mainly by simulations.

Multivariate  data can be  modeled by the sequential regression, copula models, random effects models or a combination of these techniques 
\cite{verbeke:2014}. The generalized estimating equation  type approach may not be  appropriate for missing data imputation since it does not explicitly model the within subject dependence \cite{tang:2017e}.
Formal MCMC algorithms have been developed for multivariate nonnormal data under some special cases.
 Tang \cite{tang:2018} developed MDA algorithms
for longitudinal binary and ordinal outcomes based respectively on a sequence of logistic regression and the multivariate probit model, and the latter approach is a type of Gaussian copula model.
Lee {\it et. al.} \cite{lee:2016}  proposed a  full data augmentation algorithm for the sequential  regression. It allows binary, ordinal, nominal and continuous outcomes, and models 
the binary and ordinal outcomes by the probit regression.   Lee {\it et. al.}  algorithm requires that
 the nominal outcomes  be put before other types of response variables, and therefore may not be suitable for longitudinal outcomes with  a natural order among variables.
Goldstein {\it et. al. } \cite{Goldstein:2009} described an algorithm for multivariate data with a hierarchical 
structure (e.g. repeated measures data at different visits nested within individuals) through the Gaussian copula-based random effects model, and 
it allows binary, ordinal, nominal and continuous outcomes. 

One main purpose of this article is to describe a MDA algorithm  for  multivariate nonnormal data  on basis of a sequence of regression models.
Section \ref{glmmar} presents the algorithm when the models used contain only the generalized linear models (GLM) such as the
linear regression for normal outcomes, logistic regression for binary and nominal outcomes, proportional odds model for ordinal outcomes, Poisson regression and negative binomial regression for count data,  or a mixture of these  models.
There is no restriction on the order of the response variables.
The MDA algorithms of Tang \cite{2016:tang, tang:2018} are special cases of the  proposed algorithm  when the data contain only one type of response variable (continuous, binary, or ordinal). 
 Section \ref{skewreg} extends the algorithm to incorporate the skew-normal and skew-t regressions for nonnormal continuous outcomes, and
 discusses the potential extension to include other types of regression models.

In Section \ref{mnarimp}, we  apply the proposed MDA algorithm to the controlled pattern imputations for  sensitivity analyses in longitudinal clinical trials.
The  missing at random (MAR) based analysis assumes that after treatment discontinuation, patients still have 
the same statistical behaviors as otherwise similar subjects who remain  in the trial. The MAR mechanism is unrealistic particularly if the early discontinuation is due to lack of efficacy or safety issues.
The regulatory  guidelines \cite{chmp:2010, ich:2017} and a  FDA-mandated panel report from the National Research Council  \cite{NRC:2010} recommend  sensitivity analysis
under missing not at random (MNAR) in the sense that the response profiles for subjects who withdraw are systematically different from those  who remain  on the treatment.
The controlled pattern imputations, pioneered by Little and Yau \cite{little:1996},
assume that the treatment effect reduces or disappears after the treatment discontinuation by taking into account of the  treatment actually received after 
dropout \cite{2013:mallinckrodta, 2013:carpenter, 2016:tangb,2016:tang,  tang:2018}. These methods have become increasingly popular
in clinical trials because the underlying MNAR assumption is clinically plausible  and easy to interpret.

Section \ref{fcsimp} describes a heuristic approach to implement  the controlled imputation.  
We employ FCS  to impute the intermediate missing data to create a monotone missing pattern. The missing data after withdrawal are then  imputed
according to the assumed  MNAR  mechanisms.
The proposed MDA and FCS imputation algorithms  are illustrated by one simulation study   in Section \ref{simsec}, and
by the analysis of two real trials in Section  \ref{realanalysis}.

Throughout the article, we use the following notations.
Let $\mathcal{G}(a,b)$ denote a gamma distribution with shape  $a$, rate $b$ and mean $a/b$.
Let $N(\mu,\sigma^2)$ be the normal distribution, and $N^+(\mu,\sigma^2)$ the positive normal distribution (i.e. normal distribution left truncated by $0$).
Let  $t(\mu,\sigma^2,\nu)$ be the t distribution with mean $\mu$, scale  $\sigma^2$, and  $\nu$ degrees of freedom (d.f.), and
$t^+(\mu,\sigma^2,\nu)$ the positive t distribution. 
Let $t_{\nu}(\cdot)$ and $T_{\nu}(\cdot)$ denote respectively the probability density function (PDF) and  cumulative distribution function (CDF) of the standard Student's t distribution with $\nu$ d.f.

\section{MDA algorithm}\label{seqregressionsec}

Let $\yv_i=(y_{i1},\ldots,y_{ip})'$ denote the $p$ response variables of interest, and $\xv_i=(x_{i1},\ldots,x_{iq})'$ the covariates  ($x_{i1}\equiv 1$ if the model contains an intercept) for subject $i=1,\ldots,n$.
We assume the covariates are fully observed.  If a covariate contains missing values, it can be treated as a response variable.
In general,  $\yv_i$'s are partially observed. 
Let $s_i$ be the dropout pattern according to the index of the last observation for subject $i$.  We have $s_i=0$ for subjects whose responses are all missing, and
$s_i=p$ if $y_{ip}$ is observed. 

Let $\yv_{io}$, $\yv_{ic}$, $\yv_{id}$ and $\yv_{iw}$ denote respectively the observed data, intermittent missing continuous data,   intermittent missing discrete data, and 
the missing data after the last observed value for subject $i$. Let $Y_o=\{\yv_{io}: i=1,\ldots,n\}$, $Y_d=\{\yv_{id}: i=1,\ldots,n\}$, $Y_c=\{\yv_{ic}: i=1,\ldots,n\}$, and $Y_w=\{\yv_{iw}: i=1,\ldots,n\}$.
Without loss of generality, we sort the data so that subjects in pattern $s$ are arranged before subjects in pattern $t$ if $s>t$.
Let $n_j$ be the total number of subjects in patterns $j,\ldots,p$.

\subsection{MDA algorithm based on a sequence of  generalized linear models}\label{glmmar}
Suppose  the joint distribution of $\yv_i=(y_{i1},\ldots, y_{ip})'$ can be factored as $f (y_{i1},\ldots, y_{ip}) =\prod_{j=1}^p f_j(y_{ij}|\zv_{ij},\betav_j,\phi_j)$, where $\zv_{i1}=\xv_i$,
$\zv_{ij}=(\xv_{i}',y_{i1},\ldots, y_{ij-1})'$ at $j\geq 2$,  $\betav_j$ is a vector of regression coefficients, and $\phi_j$ is the   dispersion parameter (e.g. variance in linear regression). 
In practice,   the relationship between $y_j$ and $\zv_{ij}$ is usually modeled by GLM  \cite{nelder:1972,McCullagh:1989} 
\begin{equation}\label{modelj}
 f_j(y_{ij}|\zv_{ij},\betav_j,\phi_j) = \exp\left[\frac{y_{ij} \theta_j-b(\theta_j)}{a(\phi_j)} + c(y_{ij},\phi_j)\right],
\end{equation}
where $\theta_j$ is the canonical parameter. For example, the binary outcome is often analyzed by the logistic regression, and count data may be fitted by  Poisson or negative binomial regressions.  Appendix \ref{glmappendix} 
lists several commonly used GLMs and provides technical details for the MDA algorithm.
In GLM  \cite{nelder:1972,McCullagh:1989},  $y_{ij}$ has mean $\mu_{ij}=\frac{\partial b(\theta_j)}{\partial\theta_j}$
and variance $V_{ij}=\frac{\partial \mu_{ij}}{\partial\theta_j }a(\phi_j)$. 
A link function $\eta_{ij}=g(\mu_{ij})$ is used to relate  $\mu_{ij}$ to the predictor variables in $\zv_{ij}$.
For notational simplicity, we assume $y_{ij}$'s are scalar, and  $\eta_{ij}= \zv_{ij}'\betav_j= \sum_{k=1}^{q} x_{ik}\alpha_{jk} + \sum_{k=1}^{j-1} \beta_{jk} y_{ik}$, where
$\betav_j=(\alpha_{j1},\ldots,\alpha_{jq},\beta_{j1},\ldots,\beta_{j,j-1})'$.
 But  $y_{ij}$ can be a vector. For example, 
a nominal variable with $k$ levels  is typically coded as $k-1$ indicator variables. 
Furthermore,  interactions between predictors are  allowed, and there is no need to include all variables in  $\zv_{ij}$ as  predictors  in model \eqref{modelj} particularly  when the number of response variables $p$ is  large.

The  likelihood for the augmented monotone data  $(Y_o, Y_d, Y_c)$ is
 $$\mathcal{L}(\betav_1,\phi_1,\ldots,\betav_p,\phi_p| Y_o,  Y_{d}, Y_c)\propto  \prod_{j=1}^p \prod_{i=1}^{n_j} f_j(y_{ij}|\zv_{ij},\betav_j,\phi_j).$$ 
We use independent priors for $(\betav_j,\phi_j)$'s. They are also independent in the posterior distribution
\begin{equation}\label{postglm}
\pi(\betav_j,\phi_j| Y_o, Y_{d}, Y_c) \propto \pi(\betav_j,\phi_j) \prod_{i=1}^{n_j} f_j(y_{ij}|\zv_{ij},\betav_j,\phi_j).
\end{equation}
Throughout, we use  $\pi(\cdot)$ and $\pi(\cdot|\cdot)$  to denote  respectively the prior and posterior densities.

The proposed MDA algorithm (labeled as {\it A})  involves repeating the following steps until convergence
\begin{itemize}
\item[$A.1$:] Draw $(\betav_j,\phi_j)$'s  from their posterior distribution \eqref{postglm} given $Y_o$ and the current imputed $(Y_d, Y_c)$.
\item[$A.2$:] Impute intermittent missing data for subject $i=1,\ldots,n$  given  $Y_o$  and the current draw of $(\betav_j,\phi_j)$'s.
\begin{itemize} 
\item[$A.2.1$:] Impute  $\yv_{id}$ given $(\yv_{io}, \yv_{ic})$  and $(\betav_j,\phi_j)$'s.
\item[$A.2.2$:] Impute  $\yv_{ic}$ given $(\yv_{io}, \yv_{id})$  and $(\betav_j,\phi_j)$'s.
\end{itemize}
\end{itemize}
The missing data $\yv_{iw}$'s after the last observed value are imputed after the posterior samples $(\betav_j,\phi_j)$'s and $(Y_d, Y_c)$  in steps $A.1$ and $A.2$ converge to their stationary distribution \cite{2016:tang,2016:tangb}. The details
will be given in Section \ref{mnarimp}.

\subsubsection{Draw of the model parameters in Step $A.1$:}
The draw of $(\betav_j,\phi_j)$ in Step $A.1$ presents little challenge since it is identical to that in the univariate regression.  In
the linear regression,  the posterior distribution of $(\betav_j,\phi_j)$ is normal-gamma \cite{2015a:tang, 2016:tangb, 2016:tang}, and $(\betav_j,\phi_j)$ can be drawn
by the Gibbs sampler described in Appendix  \ref{normgamma}. The sampling of $\phi_j$ depends on the specific model.
 In general,  $\betav_j$ can be drawn via  Gamerman's \cite{gamerman:1997}
Metropolis-Hastings (MH) sampler or its variant. It is the Bayesian  analogue to  the iteratively reweighted least squares (IRLS) algorithm   \cite{nelder:1972,McCullagh:1989} for the maximum likelihood estimation (MLE).
We define a transformed dependent variable $y_{ij}^*$, and it is approximately normally distributed  
\begin{equation}\label{adjvardist}
y_{ij}^* = z_{ij}'\betav_j + (y_{ij}-\mu_{ij}) \frac{d\eta_{ij}}{d\mu_{ij}} \sim N[z_{ij}'\betav_j, w_{ij}(\betav_j)], \text{ where }  w_{ij}(\betav_j)=\left(\frac{d\eta_{ij}}{d\mu_{ij}}\right)^2 V_{ij}.
\end{equation}
Suppose the prior for $\betav_j$ is $N(v_j, R_j^{-1})$, and it is flat $\pi(\betav_j)\propto 1$ as $R_j\rightarrow 0$. 
Let $\Uvv(\betav_j)$ and $I(\betav_j)$ be respectively the score and Fisher information matrix for model \eqref{modelj}.
In general,  we have $\Uvv(\betav_j)= \sum_{j=1}^{n_j} z_{ij}w_{ij}^{-1}(\betav_j)  (y_{ij}-\mu_{ij}) \frac{d\eta_{ij}}{d\mu_{ij}}    $ and $I(\betav_j)=\sum_{j=1}^{n_j} z_{ij}w_{ij}^{-1}(\betav_j) z_{ij}'$. 
Let $\hat\betav_{j}= I(\betav_j)^{-1} \sum_{j=1}^{n_j} z_{ij}'w_{ij}^{-1}(\betav_j)y_{ij}^*$, and $\Sigma(\betav_j)= [\Ivv(\betav_j) + R_j^{-1}]^{-1}$. 
Gamerman \cite{gamerman:1997} uses the following proposal distribution obtained  from the approximate linear model \eqref{adjvardist}
\begin{equation}\label{adjvarpost}
\betav_{j}^* \sim N\left[\Sigma(\betav_j)\left(I(\betav_j)\hat\beta_j+ R_j v_j\right), \Sigma(\betav_j)\right].
\end{equation}

However, it is not straightforward to define the transformed variable $y_{ij}^*$ for ordinal or nominal outcomes or when there are nonlinear predictors. By noting that the IRLS algorithm
is equivalent to Fisher's score algorithm  \cite{nelder:1972,McCullagh:1989}, Tang \cite{tang:2018} proposes to sample the candidate $\betav_j^*$ from 
\begin{equation}\label{adjvarpost2}
\betav_j^* \sim N[\betav_j+\Sigma(\betav_j)( \Uvv(\betav_j)+ R_jv_j),  \Sigma(\betav_j)].
\end{equation}
The proposal distributions \eqref{adjvarpost} and \eqref{adjvarpost2} are similar especially when the prior for $\betav_j$ is noninformative. We will use the latter one.
At each MCMC iteration, a candidate $\betav_j^*$ is drawn from the proposal distribution \eqref{adjvarpost2}. We accept the move $\betav_j \rightarrow \betav_j^*$ with probability $A_{j\beta}$, and otherwise keep $\betav_j$ unchanged, where  
$\phi(\betav_j^*|\betav_j)$ is the PDF of the proposed distribution, and
$$A_{j\beta}=\min\left\{1, \frac{ \phi[\betav_j | \betav_j^*]\, \pi(\betav_j^*) \prod_{i=1}^{n_j} f_j(y_{ij}|\zv_{ij},\betav_j^*,\phi_j) }
   {\phi[\betav_j^* | \betav_j] \,\pi(\betav_j)  \prod_{i=1}^{n_j} f_j(y_{ij}|\zv_{ij},\betav_j,\phi_j) } \right\}.$$ 

If the dimension of $\betav_j$ is large, we may split $\betav_j$ into several blocks, and sample them separately using the above MH sampler.
There are possible alternative ways  to sample $\betav_j$'s. For example, for the analysis of dichotomous  and polychotomous response using the probit or ordered probit regression, one may use either 
the Gibbs sampler  through the data augmentation and  parameter expansion  (PX) techniques \cite{albert:1993, liu:1999,liu:2000,tang:2018}, or the above MH sampler.
One shall avoid using the MH within partially collapsed Gibbs (PCG) samplers \cite{vandyk:2015} if $(\betav_j,\phi_j)$ is drawn via the data augmentation technique since the stationary distribution of the Markov chain
may change. See Section  \ref{skewreg} for  further discussion. 

\subsubsection{Imputation of intermittent missing discrete outcomes in Step $A.2.1$:}
Let $\mathcal{B}_{di}$ be the set of indices for intermittent missing discrete observations, and  $h_{id}$ the index of the first missing discrete observation
for subject $i$. Let $K_j$ be the number of levels for variable $j \in \mathcal{B}_{di}$. For count data, the number of categories is infinite, and can be truncated at
 a large finite value $K_j$ at which $\Pr(y_{ij}>K_j)\approx 0$.
There are $K_{di}=\prod_{j \in \mathcal{B}_{di}} K_j$ possible combinations of  $\yv_{id}$ (denoted by $\yv_{id}^{l}$, $l=1,\ldots,K_{di}$). Set  
$\yv_{id}=\yv_{id}^{l}$ with probability $\alpha_l/\sum_{l=1}^{K_{di}} \alpha_l$, where 
$\alpha_l = \prod_{j=h_{id}}^{s_i} f(y_{ij}|\zv_{ij},\betav_j, \phi_j, \yv_{id}=\yv_{id}^{l})$.

\subsubsection{Imputation of intermittent missing continuous outcomes  in Step $A.2.2$:}\label{postconvar}
Sampling $\yv_{ic}$'s poses challenges. We focus on the  case when the minus Hessian matrix
$V_{ic}=-\sum_{j=h_{ic}}^{s_i}\frac{\partial^2 \ell_{ij}}{\partial \yv_{ic}\partial \yv_{ic}'}$ is positive definite. It holds at least for those commonly used GLMs listed  in Appendix \ref{glmappendix}
 (we will discuss  later in this section how to handle the special situation when  model \eqref{modelj} contains interactions between two intermediate missing continuous variables),
where $\ell_{ij}=\log[f(y_{ij}|\zv_{ij},\betav_j,\phi_j)]$, and $h_{ic}$ is the index of the first missing continuous observation for subject $i$.  Section \ref{skewreg}
 will briefly discuss the sampling schemes for  non-positive definite $V_{ic}$.
  
The sampling method for $\yv_{ic}$ is similar to that for $\betav_j$. 
Let $\Delta_{ic}=V_{ic}^{-1} [\sum_{j=h_{ic}}^{s_i} \frac{\partial \ell_{ij}}{\partial \yv_{ic}}]$.
 At each MCMC iteration, a candidate $\yv_{ic}^{*}$  is generated from $N[\yv_{ic}+\Delta_{ic},  V_{ic}^{-1}]$, and accepted  with probability $A_{jy}$,
where $\phi[\yv_{ic} | \yv_{ic}^{*}]$ is the PDF of the proposal distribution,  and
$A_{jy}=\min\left\{1, \frac{ \phi[\yv_{ic} | \yv_{ic}^{*}]\,  \prod_{j=h_{ci}}^{s_i} f_j(y_{ij}^*|\zv_{ij}^*,\betav_j,\phi_j) }
   {\phi[\yv_{ic}^{*} | \yv_{ic}] \,  \prod_{j=h_{ci}}^{s_i} f_j(y_{ij}|\zv_{ij},\betav_j,\phi_j) } \right\}.$

If the imputation contains only the normal linear models with the conditional mean $\text{E}(y_{ij}|y_{i1},\ldots,y_{ij-1})=  \sum_{k=1}^{q} x_{ik}\alpha_{jk} + \sum_{k=1}^{j-1} \beta_{jk} y_{ik}$, the MH sampler for $\yv_{ic}$ becomes a Gibbs sampler ($A_{jy}\equiv 1$) and 
 the proposed algorithm reduced to the MDA algorithm \cite{2015a:tang, 2016:tang} for multivariate normal data (except that the priors may be different).
For longitudinal binary or ordinal  outcomes, the above algorithm is identical to that of Tang   \cite{tang:2018}.

If model \eqref{modelj} contains interactions between two intermediate missing continuous variables for a subject, $V_{ic}$ has a complicated expression and may be non-positive definite. The missing values for this subject
can be split into few blocks (no two variables in an interaction term are in the same block), and imputed separately using the above MH sampler.

\subsection{Extension to incorporate the skew-t / skew-normal regression or other  models}\label{skewreg} 
\subsubsection{Skew-t / skew-normal regression}
In Section \ref{glmmar}, the continuous outcome is modeled by the normal linear regression. For nonnormal continuous data, one simple way is to apply some transformation 
to make the data approximately normally distributed \cite{white:2011}. However, such transformation may not always exist. Furthermore,
transformation may distort the relationship between variables \cite{lee:2017}, or make the result  difficult to interpret. 
We model the nonnormal continuous data by the skew-t or skew-normal  regression.

A continuous random variable $y$ is said to follow the skew-t distribution if  its PDF is given by \cite{azzalini:2003}
\begin{equation}\label{distat2}
f_{\mathcal{ST}}(y; \mu, \omega^2, \lambda,\nu)=\frac{2}{\omega} t_{\nu} \left( \frac{y-\mu}{\omega}\right) T_{\nu+1} \left[\lambda \frac{y-\mu}{\omega}\sqrt{\frac{\nu+1}{\nu+\frac{(y-\mu)^2}{\omega^2}}}\right],
\end{equation}
where $\mu$ is the location parameter, $\omega^2$ is the scale parameter, $\lambda$ is the skewness parameter, and $\nu$ is d.f.
It would be easier to develop the Gibbs sampling scheme  on basis of the stochastic representation for the skew-t random variable 
\begin{equation}\label{skewtdist0}
y = \mu + \frac{1}{\sqrt{d}}  [\psi \wc^* + \epsilon] =  \mu +\psi \wc+ \frac{1}{\sqrt{d}}  \epsilon,
\end{equation}
where $d \sim \mathcal{G}(\nu/2,\nu/2)$, $\wc^*\sim N(0,1)$, $\wc=\wc^*/\sqrt{d} \sim N(0,1/d)$,  and $\epsilon\sim N(0,1/\gamma)$. The parameters in equations \eqref{distat2}   
and \eqref{skewtdist0} satisfy that $\gamma=(1+\lambda^2)/\omega^2$, 
$\psi= \lambda /\sqrt{\gamma}$, and  that $\omega^2=1/\gamma +\psi^2$, $\lambda=\psi\sqrt{\gamma}$. We denote the skew-t  distribution by $\mathcal{ST}(\mu,\omega^2,\lambda, \nu)$
or  ${ST}(\mu, \psi,  \gamma,\nu)$.

The skew-t distribution  becomes the skew-normal distribution  \cite{AZZALINI:1985} if we set $\nu\equiv \infty$ (i.e. $d \equiv 1$).  The skew-normal distribution is suitable only for mildly or moderately nonnormal data since its maximum  skewness is $0.995$, and the maximum kurtosis is $0.869$  \cite{AZZALINI:1985}. 
The skew-t distribution  reduces to  the Student's t distribution at $\psi=\lambda \equiv 0$, and it can not model skewed data. The skew-t distribution allows a higher degree of skewness and/or  kurtosis \cite{azzalini:2003}.
 
In the sequential regression, we  model the nonnormal continuous outcome $y_{ij}$ by ${ST}(\zv_{ij}'\betav_j,\psi_j, \gamma_j,\nu_j)$
\begin{equation}\label{skewtdist}
y_{ij} = \zv_{ij}'\betav_j  +\psi_j \wc_{ij}+ \frac{1}{\sqrt{d_{ij}}}  \epsilon_{ij} = \zv_{ij}^{*'} \betav_j^* + \frac{1}{\sqrt{d_{ij}}}  \epsilon_{ij} ,
\end{equation}
where  $d_{ij}\sim \mathcal{G}(\nu/2,\nu/2)$,  $\wc_{ij}\sim N(0,1/d_{ij})$, $\zv_{ij}^*=(\wc_{ij},\zv_{ij}')'$, and $\betav_j^*=(\psi_j,\betav_j')'$.

\subsubsection{The prior}
In the skew-normal and skew-t regressions, there is  a non-negligible chance that the likelihood function is a monotone function of $\lambda_j =\psi_j\sqrt{\gamma_j} $  (when other parameters are fixed), and 
 the  Bayes estimate of $\lambda_j$ can be infinite if a diffuse prior is used \cite{liseo:2006, branco:2013}. The problem can be resolved by using the Jeffreys prior \cite{liseo:2006}. 
This prior has no closed-form expression, but can be well approximated by the Student's t density \cite{bayes:2007,branco:2013}. We adopt this Student's t prior 
 $\lambda_j =\psi_j\sqrt{\gamma_j} \sim t(0,\pi^2/4,1/2)$, and it can be expressed as a hierarchical prior 
$$ d_{\psi_j}\sim \mathcal{G}\left(\frac{1}{4},\frac{1}{4}\right) \text{ and } \psi_j|\gamma_j, d_{\psi_j} \sim N\left(0, \frac{\pi^2}{4d_{\psi_j}\gamma_j}\right).$$

We put a half t prior \cite{gelman:2006, huang:2013} on  $\sigma_j=\sqrt{1/\gamma_j}$ with PDF $\pi(\sigma_j)\propto [1+(\sigma_j/a_0)^2/n_0]^{-(n_0+1)/2}$. The prior can be equivalently expressed as
 a hierarchical prior 
\begin{equation}\label{gammaprior} 
 \rho_j\sim \mathcal{G}\left(\frac{1}{2},\frac{1}{a_0^2}\right)  \text{ and } \gamma_j \sim \mathcal{G}\left(\frac{n_0}{2}, n_0 \rho_j\right).
\end{equation}
Setting $n_0=2$ and $a_0=10^5$ leads to a highly noninformative prior \cite{huang:2013}. In the normal linear regression,
one popular prior for $\gamma_j$ is $\gamma_j\sim \mathcal{G}(\rho,\rho)$ for a small fixed  $\rho$, and it reduces to the Jeffreys prior $\pi(\gamma_j)\propto \gamma_j^{-1}$ as $\rho \rightarrow 0$.
As explained in Appendix \ref{betagammapost}, the gamma or Jeffreys   prior can be quite informative or inappropriate for  highly skewed data.

Inference about  $\nu_j$ also poses challenges \cite{fernandez:1999,fonseca:2008}.  As $\nu_j\rightarrow \infty$, the skew-t regression converges to the skew-normal regression, and
the estimate of $\nu_j$ can be quite sensitive to the shape of  the  prior density of $\nu_j$. 
We use the penalized complexity (PC) prior \cite{simpson:2017}  because it shows good performance in the Student's t regression in simulation. 
 It is obtained  through penalizing the complexity between the t and normal distributions, and is invariant to reparameterization. 
The PC prior density is derived in Appendix \ref{prepostnu},  which is not given by Simpson {\it et al} \cite{simpson:2017}.
In the PC prior, $\nu_j$ is bounded below by $\nu_l=2$. We also put an upper bound $\nu_m=1000$ on $\nu_j$ because the prior density can not be accurately computed  at very large $\nu_j$ due to rounding errors.
The choice of $\nu_m$  has little impact on the imputation since the skew-t density function changes little when $\nu_j>100$. Alternatively, one may use the  reference prior 
derived by  Fonseca {\it et al} \cite{fonseca:2008}. 

\subsubsection{MCMC algorithm}
At Step $A.1$ of algorithm $A$, we draw the model parameters using the following data augmentation technique by treating  $(\wc_{ij},d_{ij})$'s as additional parameters.
The  MCMC scheme for the skew-t regression can be easily modified for the Student's t or skew-normal regression by restricting $\psi_j\equiv0$ or  ($d_{ij}\equiv 1$, $\nu_j\equiv \infty$).
The details are given in a companion paper \cite{tang:2019}.

\begin{itemize}
\item[P1.] Update $\rho_j \sim \mathcal{G}(( n_0+1)/2, n_0\gamma_j+1/a_0^2)$
\item[P2.] Update $d_{\psi_j} \sim \mathcal{G}(3/4, 1/4+ 2\gamma_j\psi_j^2/\pi^2)$.
\item[P3.] Update $(\psi_j,\betav_j, \gamma_j)$'s from the gamma-normal distribution  \eqref{postgamma}  via Gibbs sampler described in Appendix \ref{normgamma}.
\item[P4.] Update $\nu_j$ via a random walk MH sampler. A candidate $\tilde\nu_j$ is drawn from $\log(\tilde\nu_j-\nu_l)\sim N[\log(\nu-\nu_l), c^2]$, and
 accepted with probability $\min\left\{1, \frac{(\tilde\nu_j-\nu_l) \pi(\tilde\nu_j)\prod_{i=1}^{n_j} f(y_{ij}|\zv_{ij},\betav_j,\gamma_j,\tilde\nu_j) }{(\nu_j-\nu_l) \pi(\nu_j)\prod_{i=1}^{n_j} f(y_{ij}|\zv_{ij},\betav_j,\gamma_j,\nu_j)}\right\}$.
If $\tilde\nu_j>\nu_m$, it will be automatically rejected. The tuning parameter $c$ will be adjusted to make the acceptance probability lie roughly in the range of $30-70\%$.
\item[P5.] Update $(d_{ij}, \wc_{ij})$ from their posterior distribution \eqref{postdwy1} for $i=1,\ldots,n_j$.
\item[PX1.]  Update $(d_{1j},\ldots,d_{n_jj}, \gamma_j)$ as $(gd_{1j},\ldots,gd_{n_jj}, \gamma_j/g)$, where $g$ is a random sample from Equation \eqref{postg}
\item[PX2.] Update $(\wc_{1j},\ldots,\wc_{n_jj},\psi_j) \rightarrow (h \wc_{1j},\ldots,h \wc_{n_jj},\psi_j/h)$, where
$H=h^2$ is drawn from Equation \eqref{posth}
\end{itemize}

In steps $A.2.1$ and $A.2.2$, the intermittent missing data  are imputed by conditioning on $(\wc_{ij},d_{ij})$'s.
Given $(\wc_{ij},d_{ij})$'s, the skew-t regression \eqref{skewtdist} becomes the normal linear regression, and $\yv_{ic}$'s
can still be imputed via the MH sampler described  in Section \ref{postconvar}.
As a {\it  cautious } note, it is inappropriate to impute $(\yv_{id}, \yv_{ic}$)'s on basis of the skew-t density $f(y_{ij}|\zv_{ij}'\betav_j,\psi_j,\gamma_j,\nu_j)$ by integrating out $(\wc_{ij},d_{ij})$'s  since this forms 
a PCG sampler, and $\nu_j$ is updated via a  MH sampler. The  stationary distribution of the Markov chain may change in an ordinary MH within the 
PCG sampler \cite{vandyk:2015}.


The PX technique  \cite{liu:1999,liu:2000} is used to speed up the convergence of the MDA algorithm. Omitting steps PX1 and PX2 does not affect the posterior distribution,
 but it may take more iterations for the Markov chain to reach stationarity with larger autocorrelation between posterior samples 
when the data are  heavy-tailed and/or highly skewed.  
Empirical experience indicates that inclusion of  steps PX1 and PX2 
tends to make the algorithm converge faster for highly nonnormal data, and there is no obvious gain in efficiency  if the data distribution is close to normal.

We assume that $V_{ic}$ is positive definite.  If a new regression model is employed in the imputation and it incurs a non-positive definite  $V_{ic}$,
some missing continuous values may  be imputed simultaneously using the proposed MH sampler if the corresponding minus Hessian matrix
is positive definite, and other intermittent missing continuous values may be imputed one at a time  in Step $A.2.2$. 
Several methods can be used to impute the individual missing variable:
1) adaptive Gibbs sampler of Gilks and Wild \cite{gilks:1992} for variables with  log-concave posterior density functions,
2) Gibbs sampler of Damlen {\it et al}  \cite{damlen:1999} through the introduction of auxiliary uniform random  variables,
3) random walk MH sampler.

\subsection{Controlled imputation for longitudinal clinical trials}\label{mnarimp} 
The controlled pattern imputation is often served as sensitivity analysis to assess the robustness of the conclusion obtained from 
the MAR-based analysis in clinical trials \cite{little:1996, 2013:mallinckrodta, 2013:carpenter, 2016:tangb,2016:tang,  tang:2018}.
For simplicity, we assume the trial consists of two treatment groups. Let $x_{iq}=g_i$ be the treatment status ($g_i=1$ for the experimental treatment, $0$ for control).

 The missing responses $\yv_{iw}$'s after dropout are imputed according to  some MNAR mechanisms. It is a type of pattern mixture model (PMM)
since the joint distribution of $\yv_i$  varies by the dropout pattern
\begin{equation}\label{glmpmm}
f(\yv_i|s_i=s) = \prod_{j=1}^{s} f_j(y_{ij}|\zv_{ij},\betav_j,\phi_j) \prod_{j=s+1}^p g_j(y_{ij}|\zv_{ij},\betav_j, \phi_j, \Deltav_{j}). 
\end{equation}
In PMMs, the distribution of the outcomes before dropout 
 $\prod_{j=1}^{s} f_j(y_{ij}|\zv_{ij},\betav_j,\phi_j)$ is the same as that  under MAR. It implies that the intermittent missing data are  MAR.
Under the MAR dropout mechanism, the distribution of the missing data after dropout $g_j(y_{ij}|\zv_{ij},\betav_j, \phi_j,\Deltav_{j})$ is identical to $f_j(y_{ij}|\zv_{ij},\betav_j,\phi_j)$.
In case of nonignorable dropout, the missing data distribution can be specified by modifying the 
linear predictor $\eta_{ij}=\zv_{ij}'\betav_j$, where $\Deltav_{j}$'s are the additional parameters to capture deviation from MAR, and 
assumed to be known since it can not be inferred from the observed data \cite{2016:tang}.

Below, we describe two types of controlled imputations. The control-based PMM, also called ``copy reference'' (CR), was initially proposed in the seminal work of Little and Yau \cite{little:1996},  and later studied by a number of authors
 \cite{2013:carpenter,2016:tang, 2016:tangb}.  The missing data after dropout are imputed  on an as-treated basis by taking into account of the  treatment actually received after withdrawal.
Specifically, it assumes that conditioning on the observed history, the statistical behavior of dropouts from the experimental arm is the same as that of subjects on the control treatment.
The imputation can be conducted by modifying $\eta_{ij}$ as (i.e. set  the treatment status $g_i=x_{iq}\equiv 0$ for all subjects after dropout)  
$$ \eta_{ij}= \sum_{k=1}^{q-1} x_{ik}\alpha_{jk} + \sum_{k=1}^{j-1} \beta_{jk} y_{ij} \text{ for } j>s_i.$$

In the delta-adjusted PMM, the response among subjects who discontinue the treatment may improve (e.g.  subjects who discontinue the placebo due to lack of efficacy may use other drugs available on the market)
or deteriorate (e.g.  subjects who discontinue the experimental  treatment due to safety) compared to subjects who  remain on the same treatment. For subjects in pattern $s$, the missing response can be imputed by shifting $\eta_{ij}$ for a 
pre-specified amount $\Delta_{sj_g}$
\begin{equation}\label{proptodel0}
\eta_{ij}= \sum_{k=1}^{q} x_{ik}\alpha_{jk} + \sum_{k=1}^{j-1} \beta_{jk} y_{ij} + \Delta_{sj_g} \text{ for } j>s.
\end{equation}

The popular {\it tipping point}  analysis  \cite{ich:2017,permutt:2016}  is  built on the delta-adjusted imputation. It assesses how severe the departure 
from MAR can be in order to overturn the MAR-based result.
The analysis is the most suitable when the data contain only one type of response variables.
To reduce the number of sensitivity parameters, we set $\Delta_{sj_g}=\Delta_g$  for all $j>s$, but other options are possible \cite{2016:tang}.  
The tipping point analysis is often implemented by assuming MAR  in the control arm (i.e. $\Delta_0=0$). The MI analysis is performed over a  sequence of prespecified values for $\Delta_1$
(which leads to worse response among dropouts from the experimental arm)
in order to find the tipping point $\Delta_1$  at which the statistical significance of the treatment effect is lost \cite{2013:mallinckrodta, 2016:tang}.  
The FDA statisticians    also recommend applying the adjustment in both treatment groups;
Please see Permutt \cite{permutt:2016} for details.
The MI analysis is repeated over a range of prespecified values for $(\Delta_0, \Delta_1)$
 in order to identify the region in which the treatment comparison becomes statistically insignificant.
If the insignificance region  is deemed clinically implausible, one can claim that the  analysis is robust to deviations from MAR.

In these PMMs, the joint likelihood of $(s_i, \yv_i)$ can be factored as
\begin{equation}\label{pospmm}
 \left\{\prod_{i=1}^n \Pr(s_i|\xv_i, \zeta)\right\} \left\{\mathcal{L}(\betav_1,\phi_1,\ldots,\betav_p,\phi_p| Y_o, Y_d, Y_c) \prod_{i=1}^n \prod_{j=s_i+1}^p  g(y_{ij}|\zv_{ij},\betav_j,\phi_j, \Deltav_{j})\right\}.
\end{equation}
If the parameters $\zeta$ and $\betav_j$'s are separable with independent priors, the marginal posterior distribution of
$(\betav_j,\phi_j)$'s in PMMs is identical to that under MAR. The missing data $\yv_{iw}$'s can be imputed based on the following algorithm $B$
\begin{itemize}
\item[$B.1$:] Run algorithm $A$  and collect $m$ posterior samples of $(\betav_j,\phi_j,\yv_{id},\yv_{ic})$'s after Algorithm $A$ converges.
Posterior samples may be retained at every $t$-th iteration for a large $t$ (say $t=50$) in order to achieve approximate independence between posterior samples. 
\item[$B.2$:] Impute $y_{ij}$'s ($j>s_i$) sequentially from $g(y_{ij}|\zv_{ij},\betav_j,\phi_j,\Deltav_{j})$ given the model  parameters drawn at Step $A.1$. 
\item[$B.3$:] Draw $\zeta$ from its posterior distribution. This step can be ignored if the purpose is to  impute $y_{ij}$'s.
\end{itemize}

\section{Fully conditional specification (FCS)}\label{fcsimp}
In this section, we describe an alternative approach to perform the controlled  imputations via FCS.
The FCS \cite{raghunathan:2001, buuren:2006,buuren:2007} is an imputation procedure for multivariate nonnormal data that may contain  different types of response variables. The data are imputed on a variable-by-variable basis by 
specifying a conditional model for each incomplete variable with all other variables as predictors
\begin{equation}\label{fcsmod}
  p(y_{ij} | \xv_i, y_{i1},\ldots, y_{i,j-1}, y_{i,j+1},\ldots, y_{ip},\Thetav_j) \text{ for } j=1,\ldots, p.
\end{equation} 
Each iteration step consists of successive draw of $(\Thetav_1, Y_{1m}),\ldots, (\Thetav_p, Y_{pm})$, where $Y_{jm}$ denotes all missing outcomes at visit $j$.
In FCS, $\Thetav_j$ is drawn from its posterior distribution  given the current imputed dataset,
\begin{equation}\label{fcspos}
 \pi(\Thetav_j| \xv_i, y_{i1},\ldots,y_{ip}) \propto \pi(\Thetav_j) \prod_{i: y_{ij} \text{ is observed}} p(y_{ij} | \xv_i, y_{i1},\ldots, y_{i,j-1}, y_{i,j+1},\ldots, y_{ip},\Thetav_j).
\end{equation}
The missing $y_{ij}$'s at visit $j$ are  imputed from model \eqref{fcsmod} given the current draw of $\Thetav_j$ and the current imputed missing values at all other visits.
The  FCS algorithm is similar to the traditional Gibbs sampler except that only information from subjects with observed $y_{ij}$ is used to draw
$\Thetav_j$. The FCS algorithm usually converges quickly \cite{buuren:2006,white:2011}.

It is flexible to specify the imputation model \eqref{fcsmod}, which may not be fully parametric. However, it is usually unknown to which stationary distribution
the algorithm converges  for complicated  conditional models, or such stationary distribution may not exist \cite{buuren:2006, liu:2014, chen:2015}.
As evidenced in some empirical studies \cite{buuren:2006,buuren:2007, lee:2010b, tang:2018}, FCS generally performs well under MAR despite  its theoretical weaknesses.
 MNAR imputation can be implemented in FCS by multiplying or shifting the imputed values by a constant
amount \cite{buuren:2011}, but the corresponding mechanism is hard to understand and interpret \cite{tang:2018}. 

We propose the following MNAR analysis via the FCS imputation. Firstly, the intermittent missing data are imputed via FCS under MAR. 
We then draw $(\betav_j^*,\phi_j^*)$ for $j=1,\ldots,p$ from model \eqref{modelj} given the imputed monotone dataset, and
impute  the missing data $\yv_{iw}$'s due to dropout  under the specific MAR or MNAR mechanism  described in Section \ref{mnarimp}.
A theoretical justification of the algorithm is given in Appendix \ref{fcsmnarapp}.

At each iteration,  $\Theta_j$'s and $(\betav_j^*,\phi_j^*)$'s are drawn once from their posterior distribution. A practical way is
to approximate the posterior distribution by the asymptotic normal distribution of the MLE  \cite{buuren:2006}. It can be computationally intensive to find the MLEs particularly if 
a large number of imputations are needed in order to stabilize the MI result \cite{2013:mallinckrodta, 2014d:lu, 2016:tangc}.

\section{Simulation}\label{simsec}

\begin{table}[htbp]
 \centering
\caption{Comparison of MI estimates from the probit regression of $y_{i2}$ on $\xv_i=(1,y_{i0}, g_i)'$  by simulation: \newline
  $^{(a)}$ average estimates over $H=1,000$ full datasets, where missing data are generated according to the true mechanism;\newline
$^{(b)}$ sample variance of $H=1000$ MI estimates;\newline
 $^{(c)}$ the estimates from MDA-ST and MDA-norm are the same since there is no continuous outcome in the analysis.
}\label{simres11}
\small
\begin{tabular}{l@{\extracolsep{5pt}}c@{\extracolsep{5pt}}l@{\extracolsep{5pt}} r @{\extracolsep{5pt}} r @{\extracolsep{5pt}}c@{\extracolsep{5pt}}c@{\extracolsep{5pt}}   c@{\extracolsep{5pt}}c@{\extracolsep{5pt}}c@{\extracolsep{5pt}}c@{\extracolsep{5pt}}c@{\extracolsep{5pt}}c@{\extracolsep{5pt}}c@{\extracolsep{5pt}}c@{\extracolsep{5pt}}ccc} \\\hline 
assumed & && Full$^{(a)}$&  \multicolumn{3}{c}{MDA-ST} & \multicolumn{3}{c}{MDA-norm} &     \multicolumn{3}{c}{FCS} \\ \cline{5-7}\cline{8-10}\cline{11-13}  
missing  &  include &                      & data  &    MI  &     Rubin's  & sample  &    MI  &     Rubin's  & sample   & MI  &     \multicolumn{1}{c}{Rubin's} & sample    \\  
mechanism & $y_{i1}$ &  parameter  &  estimate  & estimate  & variance & variance$^{(b)}$         & estimate  & variance & variance$^{(b)}$         &    estimate  & total & variance$^{(b)}$    \\\hline
\multicolumn{12}{c}{$y_{i1}$ is normally distributed }\\
MAR  & Yes         &         intercept&$-0.097$&$  -0.105$&$   0.015$&$   0.015$&$  -0.106$&$   0.015$&$   0.015$&$  -0.114$&$   0.015$&$   0.015$\\
                                       &&                                 $y_{i0}$ & $0.662$&$   0.659$&$   0.011$&$   0.013$&$   0.658$&$   0.011$&$   0.013$&$   0.651$&$   0.012$&$   0.013$\\ 
                                        &&                                        treatment&$0.805$&$   0.806$&$   0.034$&$   0.036$&$   0.802$&$   0.034$&$   0.036$&$   0.798$&$   0.035$&$   0.035$\\
                                                                                  
 &  NO $^{(c)}$ &          intercept&$-0.097$&&&&$  -0.263$&$   0.017$&$   0.017$&$  -0.271$&$   0.017$&$   0.017$\\
                &&                                                                       $y_{i0}$& $0.662$&&&&$   0.617$&$   0.013$&$   0.015$&$   0.599$&$   0.013$&$   0.015$\\ 
                 &&                                                                              treatment&$0.805$&&&&$   0.734$&$   0.041$&$   0.042$&$   0.730$&$   0.040$&$   0.041$\\     

 CR & YES &                intercept&$-0.100$ &$  -0.104$&$   0.015$&$   0.015$&$  -0.106$&$   0.015$&$   0.015$&$  -0.113$&$   0.015$&$   0.015$\\
                  &&                                                      $y_{i0}$&$0.648$&$   0.648$&$   0.011$&$   0.012$&$   0.647$&$   0.011$&$   0.012$&$   0.640$&$   0.012$&$   0.012$\\ 
                 &&                                                                treatment&$0.760$&$   0.759$&$   0.032$&$   0.022$&$   0.758$&$   0.032$&$   0.022$&$   0.749$&$   0.033$&$   0.021$\\
             
 &    NO$^{(c)}$ &            intercept&$-0.100$ &&&&$  -0.256$&$   0.017$&$   0.016$&$  -0.264$&$   0.017$&$   0.016$\\
                &&                                                                       $y_{i0}$&$0.648$&&&&$   0.549$&$   0.013$&$   0.014$&$   0.531$&$   0.013$&$   0.014$\\ 
                 &&                                                                              treatment&$0.760$&&&&$   0.412$&$   0.034$&$   0.012$&$   0.405$&$   0.033$&$   0.012$\\
\\
\multicolumn{12}{c}{ $y_{i1}$ follows the skew-t distribution  }\\
MAR  & Yes         &      intercept&  $-0.084$ &$  -0.081$&$   0.013$&$   0.013$&$  -0.078$&$   0.014$&$   0.013$&$  -0.099$&$   0.014$&$   0.013$\\
                  &&                                                      $y_{i0}$&$0.466$&$   0.471$&$   0.009$&$   0.009$&$   0.470$&$   0.009$&$   0.009$&$   0.466$&$   0.009$&$   0.009$\\ 
                   &&                                                            treatment&$0.582$&$   0.566$&$   0.028$&$   0.029$&$   0.562$&$   0.029$&$   0.029$&$   0.561$&$   0.030$&$   0.029$\\

 &  NO$^{(c)}$&            intercept&$  -0.084$& &&&$  -0.421$&$   0.019$&$   0.019$&$  -0.421$&$   0.018$&$   0.019$\\
                                &&                                        $y_{i0}$&$0.466$&&&&$   0.437$&$   0.012$&$   0.013$&$   0.428$&$   0.012$&$   0.013$\\ 
                                 &&                                                treatment&$0.582$&&&&$   0.504$&$   0.040$&$   0.041$&$   0.510$&$   0.039$&$   0.041$\\

 CR & YES &                           intercept& $-0.080$ &$  -0.081$&$   0.013$&$   0.013$&$  -0.078$&$   0.014$&$   0.013$&$  -0.099$&$   0.014$&$   0.013$\\
                            &&                                            $y_{i0}$&$0.461$&$   0.464$&$   0.008$&$   0.008$&$   0.465$&$   0.009$&$   0.008$&$   0.461$&$   0.009$&$   0.009$\\ 
                             &&                                                   treatment&$0.547$&$   0.537$&$   0.027$&$   0.020$&$   0.535$&$   0.027$&$   0.020$&$   0.524$&$   0.028$&$   0.019$\\
                                                      
 &    NO$^{(c)}$ &         intercept& $-0.080$&&&&$  -0.416$&$   0.018$&$   0.018$&$  -0.416$&$   0.018$&$   0.018$\\
                     &&                                                   $y_{i0}$&$0.461$&&&&$   0.397$&$   0.012$&$   0.013$&$   0.388$&$   0.012$&$   0.012$\\ 
                      &&                                                           treatment&$0.547$&&&&$   0.281$&$   0.033$&$   0.013$&$   0.279$&$   0.032$&$   0.013$\\

\hline
\end{tabular} 
\end{table}

In this simulation, we assess whether the use of intermediate outcomes can improve the MI inference in the controlled imputation, and   
compare the performance of the  normal versus skew-t  regressions in imputing continuous outcomes. 
The following priors are used in all  numerical examples.   In the skew-t regression, we set the prior parameter $\varrho=p_0/d(\nu_0)$ on basis of the prior belief that there is a 
$p_0=70\%$ chance that $\nu_j$ is below $\nu_0=10$. Empirical experience indicates that the MI result is quite insensitive to the choice of $p_0$.
For the normal linear regression, we use the prior $\pi(\betav_j,\gamma_j)\propto \gamma_j^{-1}$. In other regressions, the prior is $\pi(\betav_j) \sim N(\zerov, R_j^{-1})$,
where $R_j=\text{diag}(10^{-8},\ldots,10^{-8})$. In the imputation algorithm, the binary outcomes are modeled by the logistic regression.

Two scenarios are considered. 
In scenario $1$, we simulate $H=1,000$ datasets of  size $n=300$ ($150$ subjects per arm)  from the following model: 
$$y_{i0}\sim N(0,1), \, y_{i1}|y_{i0},g_i \sim N(0.5 + 0.5y_{i0}+g_i,1), \Pr(y_{i2}=1 |y_{i0},y_{i1},g_i) =\Phi (-0.5+0.25y_{i0}+0.8y_{i1}),$$
where $\Phi(\cdot)$ is the CDF of $N(0,1)$.
We can view $y_{i1}$  as a surrogate for  $y_{i2}$ in the sense that 
the treatment effect on $y_{i2}$ is totally mediated through $y_{i1}$.
Pattern is generated according to $\Pr(s_i=0)=\text{expit} (0.3y_{i0}-3)$, 
 and $\Pr(s_i=1|s_i\geq 1)=\text{expit}(0.3y_{i0} + y_{i1}-2)$, where $\text{expit}(x) = \frac{\exp(x)}{1 + \exp(x)}$.
The proportions of subjects in patterns $0$, $1$ and $2$  are approximately $(4.88\%, 30.53\%, 64.59\%)$. 
Intermittent missing data are generated by setting  $y_{ij}$  ($1\leq j< s$)  to be missing with a $20\%$ chance among pattern $s$. The baseline $y_{i0}$ is observed in all subjects.
Scenario $2$ is similar to scenario $1$ except that $y_{i1}$ is generated from a skew-t distribution with parameters 
 $\mu_{i1}=0.5-2\sqrt{2/\pi} + 0.5 y_{i0}+g_i$, $\psi=2$, $\gamma=1$ and $\nu=10$.

We assess the treatment effect on $y_{i2}$.
Each simulated dataset is imputed using all observed information under both MAR and CR  by the MDA and FCS algorithms.
 In MDA,  $y_{i1}$ is assumed to be either normally distributed 
(labeled as ``MDA-norm'') or skew-t distributed (labeled as ``MDA-ST'').
  We set $\xv_i=(1,y_{i0}, g_i)'$.
In FCS,  $m=100$ datasets are imputed after a burn-in period of $200$ iterations. 
In MDA, $m=100$ posterior samples are collected  every $50$th iteration after a burn-in period of $5,000$ iterations.
Each imputed dataset is analyzed by fitting  a probit regression of $y_{i2}$ on $\xv_i=(1,y_{i0}, g_i)'$. The results from the $m$ imputed datasets are combined for inference 
via Rubin's rule \cite{rubin:1987}. The whole analyses are then repeated by excluding $y_{i1}$ in the imputation.

The results are reported in Table \ref{simres11}. 
The full  data estimate is the average of $H = 1000$ complete data estimates, where the missing data after dropout are generated according to the true 
mechanism at the true parameter values. 
Compared to FCS, both MDA-ST and MDA-norm yield slightly better results  in the sense that the 
MI estimates  are closer to the full data estimate, and have smaller MI  variance.
When $y_{i1}$ is normally distributed, the performance of MDA-ST is almost as good as MDA-norm.
MDA-ST exhibits some improvement over MDA-norm when $y_{i1}$ is skew-t distributed. 

The CR  approach yields more conservative  treatment effect estimates and slightly smaller MI variance estimates than the MAR-based analysis when $y_{i1}$ is included in the imputation.
The differences in the MI treatment effect  and variance estimates between the MAR and CR approaches become more pronounced when $y_{i1}$ is excluded from the analysis. 
Rubin's MI variance estimates are close to the sampling variance under MAR. 
In the CR approach, Rubin's rule  overestimates the sampling variance of the treatment effect,
but not the sampling variances for the intercept and the coefficient of $y_{i0}$. 
For example,  the sample variance for the H = 1000 treatment effect estimates under CR is  $0.012$, but 
 Rubin's variance averaged over the $H=1000$ replications is $0.034$ when $y_{i1}$ is normally distributed, and excluded from the MDA-norm imputation. 
The bias in Rubin's variance estimator is due to the uncongeniality between the imputation and analysis models. Similar phenomena are observed in the analysis of
longitudinal continuous \cite{2016:tangc}  and binary \cite{tang:2018} outcomes.  

The  MI estimates are close to the full  data estimates if we include $y_{i1}$ in the imputation under both  MAR and CR.
After we exclude $y_{i1}$ from the imputation,  the treatment effect estimate reduces and Rubin's variance estimate increases under both MAR and CR.
This is particularly obvious in the CR approach. For example,  the  treatment effect estimate under CR   is $0.758$ when $y_{i1}$ is normally distributed and included in the MDA-norm imputation, 
compared to  $0.412$ if $y_{i1}$ is excluded from the analysis. 
This example indicates that excluding important outcomes in the imputation may increase the bias  and variance in the parameter estimation.

\section{Real data examples}\label{realanalysis}

\subsection{Analysis of an antidepressant trial}

The antidepressant clinical trial has been analyzed by several authors  \cite{2013:mallinckrodta, 2015a:tang, 2016:tangb, 2016:tang} to illustrate the missing data methodologies.  
The Hamilton 17-item  rating scale for depression (HAMD-17) 
is collected at baseline and weeks 1, 2, 4 and 6. The dataset consists of 84 subjects on the experimental treatment and 88  subjects on placebo. The dropout
rate is $24\%$ ($20/84$) in the experimental arm and $26\%$ ($23/88$) in the placebo arm.

The endpoint could be either  a binary outcome defined as a $50\%$ improvement in HAMD-17  from baseline, or a continuous outcome defined as the change 
from baseline in HAMD-17. This binary endpoint is 
clinically relevant in assessing the efficacy of an antidepressant \cite{EMA:2011}. Suppose  it is of interest to  estimate
the effect of the test product   compared to placebo on the HAMD-17 improvement rate at week $6$. 
For illustrative purposes, the data at week 1, 4, 6 ($y_{i1}$, $y_{i3}$ and $y_{i4}$)  are analyzed as  binary endpoints, and the data at week 2  ($y_{i2}$)
are treated as an ``intermediate'' continuous outcome. 

The data are imputed under both MAR and CR    in two different strategies. 
In one strategy, all observed data at baseline and four post-baseline visits are employed to impute the missing responses, and $\xv_i=(1,y_{i0},g_i)$.
In the second strategy, $y_{i2}$ is excluded from the imputation.
We impute $10,000$ datasets using   MDA-ST ($y_{i2}$ is assumed to be skew-t distributed), MDA-norm ($y_{i2}$ is assumed to be normally distributed) and FCS.
The imputed data at week $6$ are analyzed by the logistic regression. 
In MDA, $10,000$ datasets are imputed from every 100th iteration after a burn-in period of $100,000$ iterations.  The convergence of the Markov chain is evidenced by the trace plots and autocorrelation function  plots.
The burn in period is set to be long enough. It takes a little more time (say $<30$ minutes) to run the analysis, but there is less concern about the convergence issue. 
This might be recommended in the analysis of pharmaceutical trials, where the  analysis is  prespecified, and may not be actually conducted by a statistician.
In FCS, $10,000$ datasets are imputed after a burn-in period of $200$ iterations.
 A large number of imputations are needed to stabilize the MI results \cite{2014d:lu, 2016:tangc}.  

 \begin{figure}[htbp]
\centering
\includegraphics[scale=0.75]{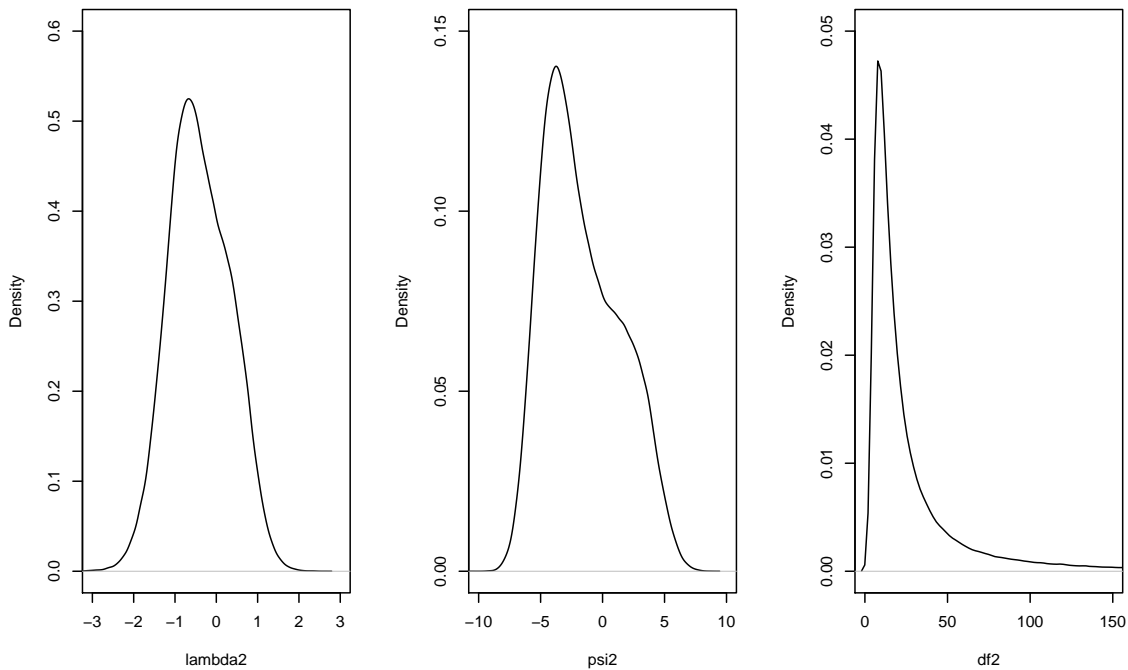}
\caption{Marginal posterior density of $\lambda_2=\psi_2\sqrt{\gamma_2}$, $\psi_2$ and $\nu_2$ (df2) from the MDA-ST  algorithm  in the analysis of an antidepressant trial: 
(a) $\lambda_2$: posterior mean $-0.397$ median $-0.442$;
(b) $\psi_2$: posterior mean $-1.607$, median $-2.155$;
(c) $\nu_2$: posterior mean $39.599$, median  $ 16.040$.
}\label{tipdep}
\end{figure}

Figure \ref{tipdep} plots the posterior density for $\lambda_2=\psi_2\sqrt{\gamma_2}$, $\psi_2$ and $\nu_2$ in the MAD-ST algorithm when $y_{i2}$ is included in the imputation. As the median $\lambda_2$ is close to $0$, and the median $\nu_2$ is $16.04$, the conditional distribution of $y_{i2}$ given $(y_{i0}, y_{i1}, g_i)$ deviates  only mildly from normality.
As displayed in Table \ref{dep10000}, MDA-ST, MDA-norm and FCS yield quite similar results. Compared to the analyses that employ $y_{i2}$ in the imputation, excluding $y_{i2}$   leads to 
 larger variance of the estimated treatment effect under  MAR, and  smaller  treatment effect estimates and larger variance under CR. 
In this example, Rubin's variance estimates under CR is  close to that under MAR.

\begin{table}[htbp]
 \centering
\caption{Estimated treatment effect on the response rate   defined as a $50\%$ improvement in HAMD-17 total score from  baseline to week $6$ in an antidepressant trial: 
 $^{(a)}$  MDA-ST and MDA-norm yield the same result since there is no continuous outcome in the analysis after excluding $y_{i2}$.
}\label{dep10000}
\begin{tabular}{l@{\extracolsep{5pt}}c@{\extracolsep{5pt}}c@{\extracolsep{5pt}}c@{\extracolsep{5pt}}c@{\extracolsep{5pt}}c@{\extracolsep{5pt}}c@{\extracolsep{5pt}}c@{\extracolsep{5pt}}c@{\extracolsep{5pt}}c@{\extracolsep{5pt}}c@{\extracolsep{5pt}}c@{\extracolsep{5pt}}c@{\extracolsep{5pt}}cccc} \\\hline 
assumed & & \multicolumn{3}{c}{MDA-ST} & \multicolumn{3}{c}{MDA-norm} &    \multicolumn{3}{c}{FCS} \\ \cline{3-5}\cline{6-8} \cline{9-11}  
missing   &    Include   &                       MI  &    Rubin's &       &        MI  &    Rubin's &  & MI  &     Rubin's  &    \\  
mechanism & $y_{i2}$ &   estimate  & variance  & t         & estimate  & variance  & t         &   estimate  & variance& t     \\\hline
  MAR&YES&$0.619$ & $0.354$ & $1.750$& $0.614$&$0.353$&$1.737$&$0.616$&$0.356$&$1.731$\\
   &NO$^{(a)}$&  & & & $0.616$&$0.365$&$1.688$&$0.612$&$0.365$&$1.677$\\ 

  CR&YES&$0.549$ & $0.348$ & $1.576$ & $0.545$&$0.347$&$1.569$&$0.545$&$0.349$&$1.561$\\ 
   &NO$^{(a)}$& & & & $0.511$&$0.354$&$1.442$&$0.509$&$0.355$&$1.434$\\ 
\hline
\end{tabular} 
\end{table}

\subsection{Analysis of the NIMH schizophrenia trial}\label{realnimh}
 We revisit    the National Institute of Mental Health (NIMH) Schizophrenia Collaborative  study analyzed by Tang \cite{tang:2018}. 
The dataset  contains $108$  subjects on placebo, and $329$ subjects on the anti-psychotic treatments.
Item 79 (severity of illness) of the Inpatient Multidimensional Psychiatric Scale  (IMPS) is collected at baseline and week 1, 3 and 6, 
and analyzed as a  binary outcome ({\it 1: normal to mildly ill, 2= moderately to extremely ill}). The dropout rate is 
about $35.2\%$ in the placebo arm,  and $19.4\%$ in the experimental  arm. In addition,
  $21$ subjects have intermittent missing data. Baseline $y_{i0}$ is not included as a covariate since about $98.6\%$ subjects are {\it moderately to extremely ill} at baseline. 

Tang  \cite{tang:2018} estimates the MI treatment effect  under the MAR, CR and delta-adjusted imputation using the MDA algorithm. We perform similar analyses 
using the FCS algorithm. We impute $10,000$ datasets after a burn-in period of $200$ iterations.  
Each imputed dataset is analyzed by the logistic regression at week $6$. 
As displayed in Table \ref{imps10000}, the results from FCS and MDA are similar (the MDA result is reproduced with $m=10,000$ imputations).

\begin{figure}[htbp]
\begin{subfigure}[b]{0.5\textwidth}
\centering
\includegraphics[scale=0.75]{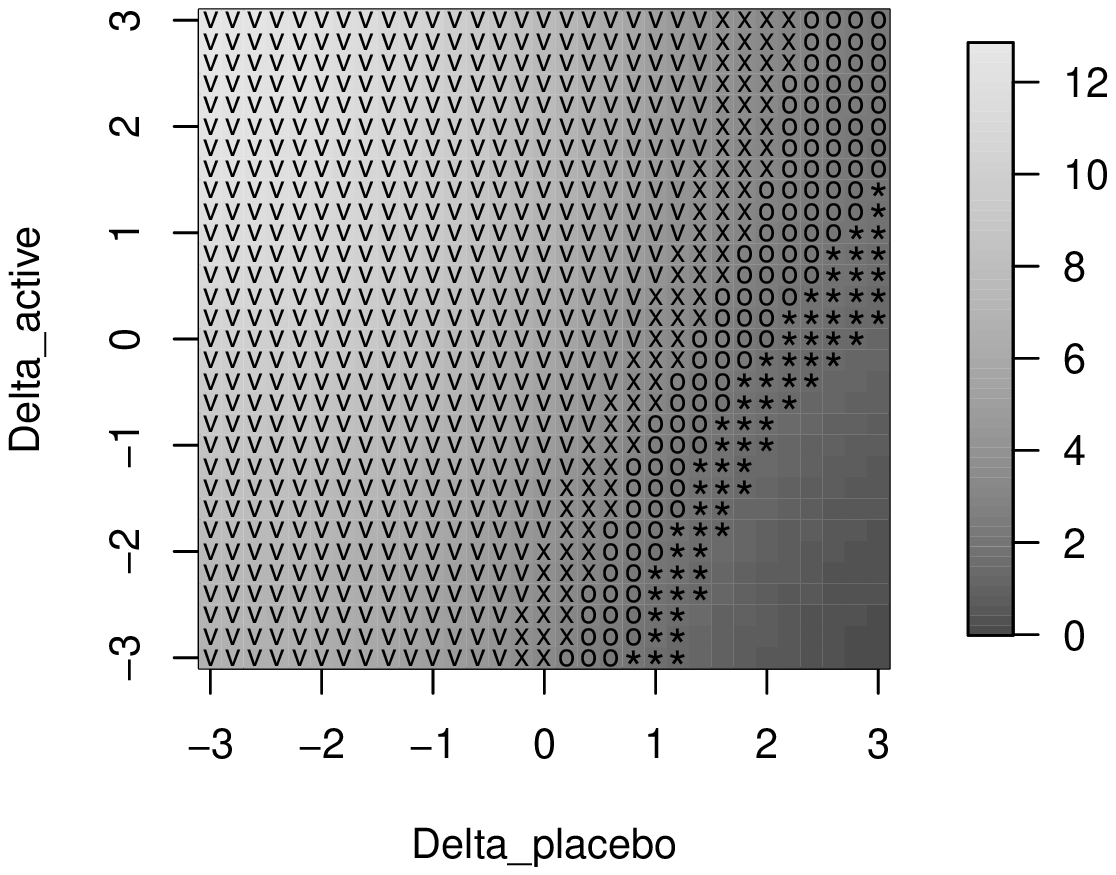}
\caption{MDA}
\end{subfigure} 
\hfill
\begin{subfigure}[b]{0.5\textwidth}
\centering
\includegraphics[scale=0.75]{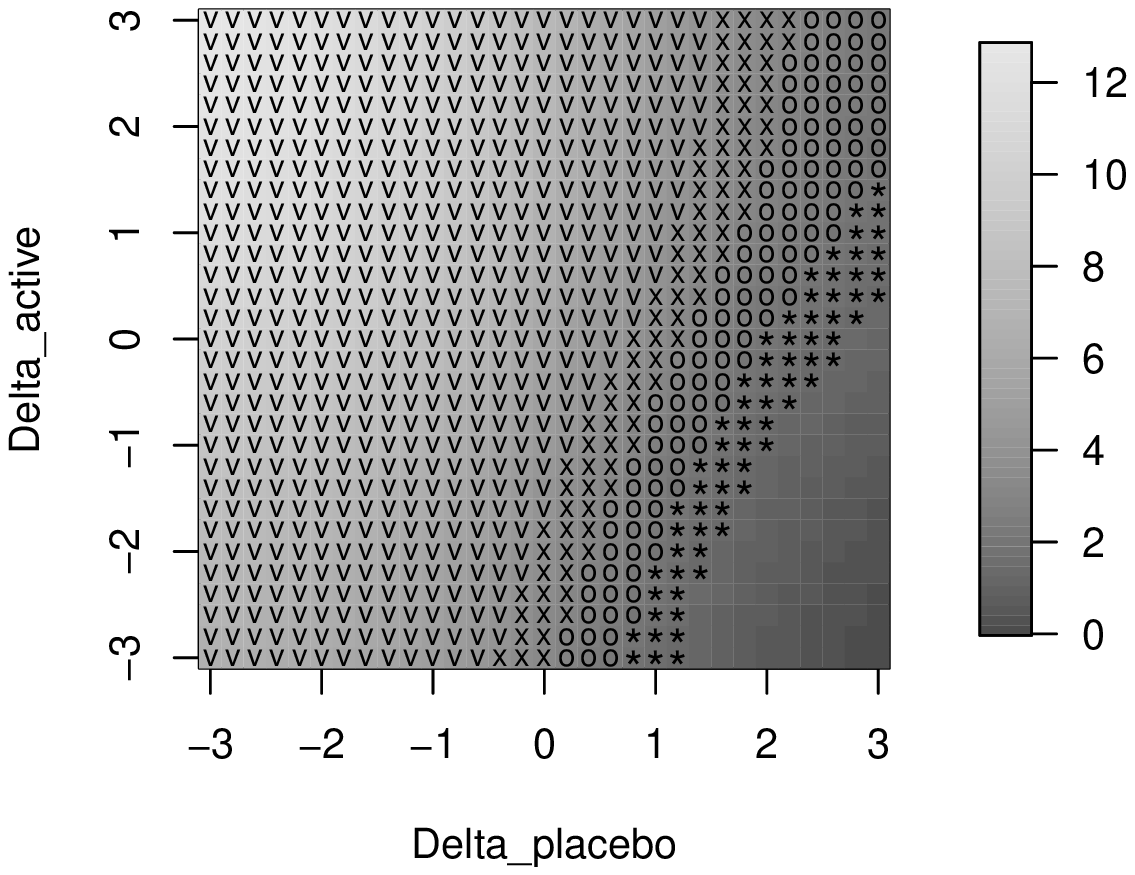}
\caption{FCS}
\end{subfigure}
\caption{Plot of $-\log_{10}(\text{pvalue})$ in the  tipping point analysis of the NIMH trial with delta adjustment in both arms: the symbols indicate the range of pvalue: `v' pvalue$<0.0001$, `x' pvalue$<0.001$, `o' pvalue $<0.01$, `*' pvalue$<0.05$}\label{tipnimh}
\end{figure}

As pointed out by Tang \cite{tang:2018}, the tipping point  does not exist if we assume MAR in the placebo arm since  the treatment comparison is still significant
when we set all missing responses  in the experimental arm  to  the worst values. 
We perform the tipping point analysis with delta adjustment in both arms. Figure \ref{tipnimh} displays the results. MDA and FCS algorithms yield very similar results.
The treatment effect at week $6$ becomes insignificant only in a small region where the odds of being ``normal to mildly ill'' among the dropouts from the experimental arm
decrease compared to subjects who remain on the experimental treatment, while the odds  of being ``normal to mildly ill''  among dropouts in the placebo arm increase compared to
subjects who remain on the placebo.

\begin{table}[htbp]
 \centering
 \def\~{\hphantom{0}}
 \begin{minipage}{160mm}
\caption{Estimated treatment effects at week $6$ and associated Rubin's variance for the NIMH Schizophrenia trial: $^{(a)}$ an adjustment of $\Delta_1=-1$ is applied to the log odds at all visits after dropout
in the experimental arm.
}\label{imps10000}
\begin{tabular}{lccccc@{\extracolsep{5pt}}c@{}rrrcrrrrrrrrrrr} \\\hline 
assumed &  \multicolumn{5}{c}{MDA} &    \multicolumn{5}{c}{FCS} \\ \cline{2-6}\cline{7-11}  
missing          &                   MI  &     \multicolumn{3}{c}{Rubin's variance} &  & MI  &     \multicolumn{3}{c}{Rubin's variance} &    \\  \cline{3-5}\cline{8-10}
mechanism &    estimate & between & within & total & t         &   estimate & between & within & total & t     \\\hline
         MAR&     $   1.417$&$ 0.024$&$ 0.060$&$ 0.084$&$ 4.886$        & $1.407$ & $0.025$ & $0.060$ & $0.085$ & $4.825$\\  
      CR&    $   1.227$&$ 0.019$&$ 0.060$&$ 0.079$&$ 4.378$        & $1.219$ & $0.020$ & $0.060$ & $0.079$ & $4.332$\\
      Delta$^{(a)}$  & $   1.259$&$ 0.024$&$ 0.060$&$ 0.084$&$ 4.344$ & $1.246$ & $0.025$ & $0.060$ & $0.085$ & $4.279$\\
\hline
\end{tabular} 
\end{minipage}
\vspace*{6pt}
\end{table}

\section{Discussion}
We develop an efficient MDA algorithm for the imputation of multivariate nonnormal data fitted by a sequence of GLMs,  skew-normal regression  and/or  skew-t regression. 
The algorithm can handle different variable types and nonnormal continuous outcomes. Its extension  to include other   models is discussed.
We apply the algorithm to the controlled imputations for the sensitivity analysis of longitudinal clinical trials. 
Due to the computational resource constraint, only one simulation study is conducted.
It demonstrates that the inclusion of important intermediate outcomes in the imputation can reduce the bias and improve the precision in estimating the treatment effect.

We also describe a heuristic approach to implement the controlled imputation via FCS.
While it is flexible to specify the conditional distribution for each individual variable given all other variables,
a theoretical weakness of FCS is that there might not exist a joint stationary distribution that is consistent with these conditional distributions \cite{buuren:2007,liu:2014,chen:2015,seaman:2016}. 
The result may be affected
by the order in which the variables are imputed \cite{chen:2015}. 
 It is unclear under what situations FCS works well, and its performance is mainly evaluated by simulations. 
The FCS  can be slightly less efficient than the MCMC-based method \cite{white:2011,lee:2016, seaman:2016}, and this is also observed in our numerical examples. 

In the CR approach,  the missing data after dropout are imputed by using the observed outcomes as predictors, and the treatment benefit obtained prior to dropout
will not disappear over a short  period of time after dropout \cite{2016:tang}. The CR assumption may not be appropriate for the situation where all the 
benefit from the treatment is gone immediately after treatment discontinuation. There are many potential ways to 
assume how the disease progresses after dropout based on the exposure-response 
relationship and/or dropout reasons. The  MDA algorithm is suitable for any PMMs that assume the same observed data distribution as that under MAR \cite{2016:tang}.

A novel MCMC algorithm is proposed for  univariate skew-t and skew-normal regressions. For  skewed and/or fat-tailed longitudinal data,
the sequential regression  introduces $p$ pairs of latent variables $( \wc_{ij},d_{ij})$'s per subject. 
In a companion paper  \cite{tang:2019}, we describe a MDA algorithm for multivariate skew-t and skew-normal  regressions. 
 The multivariate model  is more parsimonious, and  the latent variables $(\wc_{i},d_{i})$ are shared by all observations within a subject.
The  skew-t and skew-normal regressions can also be  incorporated  into FCS to handle nonnormal continuous outcomes. 

There are several potential advantages to use the  skew-t  regression to impute nonnormal continuous data. Firstly, 
the  inference is more robust to extreme outliers \cite{tang:2019}. Secondly, it may improve the precision of the treatment effect estimate, and
this is evidenced in our simulation. 
Previous studies \cite{hippel:2013, lee:2017} indicate that imputing skewed continuous data using a normal model performs well in estimating the linear regression 
coefficients (this can be justified by   Tang's \cite{2016:tangc} theoretical result that the MI and likelihood-based inferences are asymptotically equivalent
 for multivariate continuous outcomes  under MAR), but does a poor job of estimating the shape parameters such as percentiles and skewness coefficients \cite{hippel:2013}. 
We expect that the performance may be improved by using the nonnormal imputation model. 

The proposed imputation procedure has some limitations. Firstly, it assumes the intermittent missing data are MAR. In general, the assumption is reasonable since the intermittent missingness is often due to reasons (e.g. scheduling difficulty) 
unrelated to the patients' health conditions, or can be predicted given the observed outcomes.  
 In a well-conducted trial,  typically only a small proportion of patients have missing data before dropout, and the MAR assumption is not expected to have 
a big impact on the analysis result  if the intermittent missing data are MNAR \cite{schafer:1997,2016:tang}. However, the inference can be misleading if there is a large amount of nonignorable intermittent missing data. 
Secondly,  the approach is fully parametric, and its performance under model misspecification requires further investigation.
 Semiparametric  techniques may be incorporated into the imputation procedure.
For example, one may fill in  the intermittent missing data using the MDA algorithm, and then employ the predictive mean matching (PMM \cite{schenker:1996})  or 
local residual draw (LRD \cite{schenker:1996}) methods to impute the missing data after dropout.
 In both PMM and LRD, the posterior samples of the model parameters from the MDA algorithm can be used directly to impute the missing values, and there is no need to regenerate them based on the augmented monotone data.
The predicted values for the incomplete variable are commonly estimated by the  normal linear regression  \cite{schenker:1996},
but they can also be obtained from the skew-normal or skew-t regression.
It is currently unclear how to efficiently impute  the intermittent missing data by PMM or LRD. 

\flushleft{ACKNOWLEDGEMENT}\\
We would like to thank the associate editor and two referees for their helpful suggestions that improve the quality of the work.

\flushleft{SUPPORTING INFORMATION}\\
Sample SAS code can be found in the online supporting information.

\appendix
\section{Appendix}
\subsection{Several commonly used generalized linear models}\label{glmappendix}
We review several GLMs commonly used to analyze continuous, binary, ordinal, nominal and count data. Technical details are provided for the  MDA algorithm.
Throughout, let $f(y_{ij}|\zv_{ij},\betav_j,\phi_j)$ denote the PDF or probability mass function (PMF) of $y_{ij}$, and $\ell_{ij}=\log[f(y_{ij}|\zv_{ij},\betav_j,\phi_j)]$.
Let  $\tilde\betav_j=(\betav_j',0,0,\ldots,0)'$ be a $(q+p)\times 1$ vector  for categorical outcomes. Thus
$\eta_{ij}=\zv_{ij}'\betav_j =(x_{i1},\ldots,x_{iq},y_{i1},\ldots,y_{ip})\tilde\betav_j$. We define
$\tilde\betav_j=(\betav_j',-1,0,\ldots,0)'$ for a continuous outcome. Then $y_{ij}-\zv_{ij}'\betav_j = -(x_{i1},\ldots,x_{iq},y_{i1},\ldots,y_{ip})\tilde\betav_j$. 
Let $\betav_{ic}$ be a subvector of $\tilde\betav_j$ containing all elements corresponding to the intermittent missing continuous values for subject $i$. 
Let  $H(\yv_{ic})=-\frac{\partial^2\ell_{ij}}{\partial\yv_{ic}\partial\yv_{ic}'}$.

\subsubsection{ Normal linear regression for continuous outcomes}
The PDF is  $f(y_{ij}|\zv_{ij},\betav_j,\gamma_j) \propto \sqrt{\gamma_j} \exp[-\frac{\gamma_j(y_{ij}-\theta_{ij})^2}{2}]$, where $b(\theta_{ij})=\theta_{ij}^2/2$, $a(\phi_j)=1/\gamma_j$, $\theta_{ij}=\eta_{ij}=\zv_{ij}'\betav_j$. 
  We have
$$\frac{\partial\ell_{ij}}{\partial\yv_{ic}}= \gamma_j(y_{ij}-\theta_{ij})\betav_{ic}, \text{ and } H(\yv_{ic})=-\frac{\partial^2\ell_{ij}}{\partial\yv_{ic}\partial\yv_{ic}'}= \gamma_j\betav_{ic}\betav_{ic}'.$$
For the skew-t regression described in Section \ref{skewreg}, the above formulae can be modified by
replacing $\theta_{ij}$ by $\zv_{ij}'\betav_j+ \psi_j \wc_{ij}$ and $\gamma_j$ by $d_{ij}\gamma_j$.

\subsubsection{Logistic regression with logit link for binary outcomes}
We code the binary outcome as $1$ or $2$. The PMF is $f(y_{ij}|\zv_{ij},\betav_j) =\pi_{ij}^{I(y_{ij}=1)} (1-\pi_{ij})^{I(y_{ij}=2)} =\exp[ I(y_{ij}=1) \theta_{ij} -b(\theta_{ij})]$, where $a(\phi_j)=1$, $\theta_{ij}=\eta_{ij}=\zv_{ij}'\betav_j$, $\pi_{ij}=\Pr(y_{ij}=1|\zv_{ij},\betav_j)= \frac{1}{1+\exp(-\zv_{ij}'\betav_j)}$, and $b(\theta_{ij})=\log[1+\exp(\theta_{ij})]$. Then 
$$\frac{\partial\ell_{ij}}{\partial \betav_j} = [I(y_{ij}=1) - \pi_{ij}]\zv_{ij} \text{ and } I(\betav_j)=\text{E}(-\frac{\partial^2\ell_{ij}}{\partial\betav_j\partial\betav_j'})=\pi_{ij}(1-\pi_{ij}) \zv_{ij}\zv_{ij}',$$\\
$$\frac{\partial\ell_{ij}}{\partial\yv_{ic}}= [I(y_{ij}=1) - \pi_{ij}] \betav_{ic}, \text{ and } H(\yv_{ic})=-\frac{\partial^2\ell_{ij}}{\partial\yv_{ic}\partial\yv_{ic}'}= \pi_{ij}(1-\pi_{ij}) \betav_{ic}\betav_{ic}'.$$

\subsubsection{ Proportional odds models for ordinal outcomes with $K$ levels}\label{podds} We use the same notations as Tang \cite{tang:2018}.
Let $\gamma_{ij_k}= \Pr(y_{ij}\leq k|\zv_{ij})=\frac{\exp(c_{j_k} +\zv_{ij}'\betav_j)}{1+\exp(c_{j_k}+\zv_{ij}'\betav_j)}$ for $1\leq k\leq K-1$, $\gamma_{ij_0}=0$, $\gamma_{ij_K}=1$, and
$\pi_{ij_k}= \Pr(y_{ij}= k|\zv_{ij})=\gamma_{ij_k}-\gamma_{ij_{k-1}}$,
where $c_{j_1}=0$ (it is absorbed into the intercept) and $c_{j_k}=\sum_{t=2}^k \exp(d_{j_t})$ [i.e. $d_{j_k}=\log(c_{j_k}-c_{j_{k-1}})$].
Then $f(y_{ij}|\zv_{ij},\betav_j)=\prod_{k=1}^K \pi_{ij_k}^{I(y_{ij}=k)}$ and $\ell_{ij}=\sum_{k=1}^K{I(y_{ij}=k)}\log(\pi_{ij_k})$.

Tang \cite{tang:2018}  updates  $\bm{\ss}_j =(d_{j_2},\ldots,d_{j_{K-1}}, \alpha_{j1},\ldots,\alpha_{jq}, \beta_{j1},\ldots,\beta_{j,{j-1}})'$ by the MH scheme, where
$d_{j_k}^* =\exp(d_{j_k})$ at $k\leq j$,  $d_{j_k}^* =0$ if $k>j$, 
$\frac{{\partial \gamma_{ij_k}}}{\partial\bm{\ss}_j} = \gamma_{ij_k}(1-\gamma_{ij_k}) [d_{j_2}^*,\ldots,d_{j_{K-1}}^*, \zv_{ij}']'$, 
$\frac{{\partial \gamma_{ij_k}}}{\partial\bm{\ss}_j} \equiv \zerov$ at $k=0,\,K$,
 $\frac{\partial \pi_{ij_k}}{\partial\bm{\ss}_j}= \frac{\partial \gamma_{ij_k}}{\partial\bm{\ss}_j} -\frac{\partial \gamma_{ij_{k-1}}}{\partial\bm{\ss}_j}$, 
\begin{equation*}
\frac{\partial\ell_{ij}}{\partial \bm{\ss}_j}=\sum_{k=1}^K \pi_{ij_k}^{-1} \frac{ \partial \pi_{ij_k}}{\partial \bm{\ss}_j} I(y_{ij}=k) \text{ and }
         I(\bm{\ss}_j)=\text{E}(-\frac{\partial^2\ell_{ij}}{\partial\bm{\ss}_j\partial\bm{\ss}_j'})=\sum_{k=1}^K\pi_{ij_k}^{-1}\left [\frac{\partial \pi_{ij_k}}{\partial \bm{\ss}_j}\right]\left[\frac{\partial \pi_{ij_k}}{\partial \bm{\ss}_j}\right]'.
\end{equation*}

Note that $\frac{\partial \gamma_{ij_k}}{\partial\yv_{ic}} = \gamma_{ij_k}(1-\gamma_{ij_k}) \betav_{ic}$ at $k=0,\ldots,K$, and
$\frac{\partial\pi_{ij_k}}{\partial\yv_{ic}}= \frac{\partial \gamma_{ij_k}}{\partial\yv_{ic}} -\frac{\partial \gamma_{ij_{k-1}}}{\partial\yv_{ic}}= \pi_{ij_k} (1-\gamma_{ij_k}-\gamma_{ij_{k-1}})\betav_{ic}$. Thus
$$\frac{\partial\ell_{ij}}{\partial\yv_{ic}}= \sum_{k=1}^K I(y_{ij}=k)(1-\gamma_{ij_k}-\gamma_{ij_{k-1}})\betav_{ic} \text{ and }
 H(\yv_{ic})= \sum_{k=1}^K I(y_{ij}=k) [ \gamma_{ij_k}(1-\gamma_{ij_k})+\gamma_{ij_{k-1}}(1-\gamma_{ij_{k-1}})]\betav_{ic}\betav_{ic}'.$$

\subsubsection{Multinomial logistic regression  for nominal outcomes with $K$ levels}
Let $\pi_{ij_k}=\Pr(y_{ij}=k)= \frac{\exp(\zv_{ij}'\betav_{j_k})}{1+\sum_{k=1}^{K-1}\exp(\zv_{ij}'\betav_{j_k})}$ for $1\leq k\leq K-1$, and
 $\pi_{ij_K}=\Pr(y_{ij}=K)= \frac{1}{1+\sum_{k=1}^{K-1}\exp(\zv_{ij}'\betav_{j_k})}$. Let $\theta_{ij_k}= \zv_{ij}'\betav_{j_k}$ and $\betav_j=(\betav_{j_1}',\ldots,\betav_{j_{K-1}}')'$.
Then $f(y_{ij}|\zv_{ij},\betav_j) =\prod_{k=1}^K \pi_{ij_k}^{I(y_{ij}=k)} $ and $\ell_{ij} = \sum_{k=1}^{K-1} I(y_{ij}=k) \theta_{ij_k} -\log[1+\sum_{k=1}^{K-1} \exp(\theta_{ij_k})]$.\\
Let  $\piv_{ij}= (\pi_{ij_1},\ldots,\pi_{ij_{K-1}})'$ and $I_{y_{ij}}=(I(y_{ij}=1),\ldots, I(y_{ij}=K-1))'$ be a vector of indicator variables.
$$ \frac{\partial\ell_{ij}}{\partial \betav_j} =(I_{y_{ij}}-\piv_{ij}) \otimes \zv_{ij} \text{ and }
  I(\betav_j)=\text{E}(-\frac{\partial^2\ell_{ij}}{\partial\betav_j\partial\betav_j'}) = [\text{diag}(\piv_{ij})- \piv_{ij}\piv_{ij}']\otimes (\zv_{ij}\zv_{ij}'),$$
$$ \frac{\partial\ell_{ij}}{\partial\yv_{ic}} = \sum_{k=1}^{K-1}[I(y_{ij}=k)-\pi_{ij_k}]\betav_{ic_k} \text{ and }
  H(\yv_{ic})= \sum_{k=1}^{K-1}\pi_{ij_k} \betav_{ic_k}\betav_{ic_k}' - (\sum_{k=1}^{K-1}\pi_{ij_k} \betav_{ic_k})(\sum_{k=1}^{K-1}\pi_{ij_k} \betav_{ic_k})',$$
where $\betav_{ic_k}$ is a sub-vector of $(\betav_{j_k}',0,\ldots,0)$ corresponding to the intermittent missing continuous values for subject $i$. 

\subsubsection{Poisson regression for count data} The PMF is
$f(y_{ij}|\zv_{ij},\betav_j)=\frac{\mu_{ij}^{y_{ij}}\exp(-\mu_{ij})}{y_{ij}!}$, where $\theta_{ij}=\eta_{ij}=\zv_{ij}'\betav_j$, $b(\theta_{ij})=\exp(\theta_{ij})$, and $\mu_{ij}=b'(\theta_{ij})=\exp(\zv_{ij}'\betav_j)$.
We get
$$ \frac{\partial\ell_{ij}}{\partial \betav_j}= [y_{ij}- \exp(\zv_{ij}'\beta_j)]\zv_{ij} \text{ and  } I(\betav_j)=\text{E}[-\frac{\partial^2{\ell_{ij}}}{\partial \betav_j\partial \betav_j'}] = \exp(\zv_{ij}'\betav_j)\zv_{ij}\zv_{ij}',$$
$$ \frac{\partial\ell_{ij}}{\partial\yv_{ic}}= [y_{ij}- \exp(\zv_{ij}'\beta_j)]\betav_{ic} \text{ and  }
 H(\yv_{ic})=-\frac{\partial^2\ell_{ij}}{\partial\yv_{ic}\partial\yv_{ic}'} = \exp(\zv_{ij}'\betav_j)\betav_{ic}\betav_{ic}'.$$

\subsubsection{Negative binomial regression for overdispersed count data} The PMF is 
$ f(y_{ij}|\zv_{ij},\betav_j,\kappa_j)= \frac{\Gamma(y_{ij}+1/\kappa_j)}{y_{ij}!\,\Gamma(1/\kappa_j)  \kappa_j^{1/\kappa_j}}  \frac{ \mu_{ij}^{y_{ij}}}{[1/\kappa_j+ \mu_{ij}]^{y_{ij}+1/\kappa_j}}$, where  
$\mu_{ij}=\exp(\zv_{ij}'\betav_j)$. 
We have
$$\frac{\partial{\ell_{ij}}}{\partial \betav_j} =\frac{y_{ij}-\mu_{ij}}{1+\kappa_j\,\mu_{ij}}\zv_{ij} \text{ and }
I(\betav_j)=\text{E}[-\frac{\partial^2{\ell_{ij}}}{\partial \betav_j\partial \betav_j'}] =\frac{\mu_{ij}}{1+\kappa_j\,\mu_{ij}}\zv_{ij}\zv_{ij}',$$
$$ \frac{\partial\ell_{ij}}{\partial\yv_{ic}}= \frac{y_{ij}-\mu_{ij}}{1+\kappa_j\,\mu_{ij}}\betav_{ic} \text{ and  }
H(\yv_{ic})=-\frac{\partial^2\ell_{ij}}{\partial\yv_{ic}\partial\yv_{ic}'} =  \frac{(1+\kappa_j \,y_{ij})\mu_{ij}}{(1+\kappa_j\,\mu_{ij})^2}\betav_{ic}\betav_{ic}'.$$

\subsection{Posterior distributions in the skew-t regression}
The joint posterior distribution of $(\nu_j, \rho_j, \gamma_j, d_{\psi_j},\psi_j, \betav_j, d_{ij}\text{'s},\wc_{ij}\text{'s})$ is
\begin{eqnarray}\label{postskew}
\begin{aligned}
& f(\nu_j,d_{ij}\text{'s},\wc_{ij}\text{'s}, \betav_j,\psi_j,\gamma_j, d_{\psi_j}, \rho_j| Y_o,Y_d,Y_c)\\
\propto &\,
\pi(\nu_j) \pi(\rho_j) \pi(\gamma_j|\rho_j) \pi(d_{\psi_j}) \pi( \psi_j|d_{\psi_j},\gamma_j) \prod_{i=1}^{j=n_j} [ f(d_{ij}) f(\wc_{ij}|d_{ij}) f(y_{ij}| \zv_{ij},\betav_j,\gamma_j,\psi_j,d_{ij},\wc_{ij})]\\ 
\propto &\,\pi(\nu_j) [\rho_j^{\frac{1}{2}-1}\exp(-\frac{\rho_j}{a_0^2})] \, [( n_0\rho_j)^{\frac{n_0}{2}} \gamma_j^{\frac{n_0}{2}-1} \exp(-n_0\rho_j \gamma_j)] \,
  [ d_{\psi_j}^{\frac{1}{4}-1} \exp(-\frac{d_{\psi_j}}{4})] \,
\left [ \sqrt{ \frac{4 d_{\psi_j}\gamma_j}{\pi^2}} \exp(-\frac{ 4 d_{\psi_j}\gamma_j \psi_j^2}{2\pi^2})\right] \\
&\qquad\qquad\qquad \prod_{i=1}^{n_j} \left\{ d_{ij}^{\frac{\nu}{2}-1}\exp(-\frac{d_{ij}\nu}{2}) \, d_{ij}^{\frac{1}{2}} \exp(-\frac{d_{ij}\wc_{ij}^2}{2})\, \sqrt{d_{ij} \gamma_j} \exp\left[-\frac{d_{ij} \gamma_j(y_{ij}-\zv_{ij}'\betav_j-\psi_j\wc_{ij})^2}{2}\right]\right\}.
\end{aligned}
\end{eqnarray}

\subsubsection{Posterior distributions of $\rho_j$ and $d_{\psi_j}$}
By Equation \eqref{postskew}, the posterior distributions of $\rho_j$ and $d_{\psi_j}$ are both gamma
\begin{eqnarray}
f(\rho_j | Y_o,Y_d,Y_c,\gamma_j,\nu_j,  d_{\psi_j},\psi_j, \betav_j, d_{ij}\text{'s},\wc_{ij}\text{'s}) \propto \rho_j^{\frac{1+n_0}{2}-1} \exp[ -\rho_j (\frac{1}{a_0^2}+ n_0\gamma_j)], \\
 f(d_{\psi_j} |Y_o,Y_d,Y_c, \nu_j, \rho_j, \gamma_j, \psi_j, \betav_j, d_{ij}\text{'s},\wc_{ij}\text{'s}) \propto d_{\psi_j}^{\frac{3}{4}-1} \exp[ -d_{\psi_j} (\frac{1}{4}+ \frac{2\gamma_j\psi_j^2}{\pi^2})].
\end{eqnarray}

\subsubsection{Generation of normal-gamma random variables}\label{normgamma}
Suppose $(\betav,\gamma) \sim \gamma^{m/2-1}\exp(-\gamma\tilde\betav'D \tilde\betav/2)$, where  $\betav$ is a $l\times 1$ vector, $\tilde\betav =(-\betav',1)'$, and
$D$ is a $(l+1)\times (l+1)$ positive definite symmetric matrix,
Let the Cholesky decomposition of $D$ be denoted by
$D=LL'$, and $C=L^{-1}$.
Let $t_j\stackrel{i.i.d}{\sim} N(0,1)$, $t_l^2 \sim \chi_{m-l}^2$.
Let $(h_1,\ldots,h_l)'=C'(t_1,\ldots,t_{l})'$. Tang \cite{2015a:tang} shows that $(\betav,\gamma)$ can be generated   as $\gamma=h_{l}^2$ and $\betav=-(h_1,\ldots,h_{l-1})'/h_{l}$.

\subsubsection{Posterior distribution of $(\psi_j,\,$\texorpdfstring{$\betav$}{\beta}${}_j, \gamma_j)$}\label{betagammapost}
 Let  $\tilde\zv_{ij}^*=(\wc_{ij},\xv_i',y_{i1},\ldots,y_{ij})'$, $\betav_j^*=(\psi_j,\betav_j')'$, $\tilde{\betav}_j^*=(-\betav_j^{*'},1)'$,
$D_j = \sum_{i\leq n_j} d_{ij}\tilde\zv_{ij}^*\tilde\zv_{ij}^{*'}$ and $E_j=\text{diag}(  \frac{ 4 d_{\psi_j} }{\pi^2},0,\ldots,0, 2n_0\rho_j)$.
The posterior distribution of $(\betav_j^*, \gamma_j)$ is  gamma-normal, 
\begin{equation}\label{postgamma}
 f(\betav_j^*,\gamma_j|\nu_j, \rho_j,  d_{\psi_j},Y_{d},Y_c,Y_{o}, d_{ij}\text{'s}, \wc_{ij}\text{'s})
   \propto \gamma_j^{ \frac{n_j+n_0+1}{2}-1}\exp\left[-\frac{ \gamma_j  \tilde\betav_j^{*'}(D_j+E_j) \tilde\betav_j^*}{2}\right].
\end{equation}
The marginal distribution of $\gamma_j$ is gamma and the conditional distribution of $(\psi_j,\betav_j)$ given $\gamma_j$ is normal. They can be
 generated using the Gibbs sampler described in Appendix \ref{normgamma}.

For $\gamma_j$, we prefer the prior specified in  Equation \eqref{gammaprior}.  Below we explain why we don't use the gamma prior  $\gamma_j\sim \mathcal{G}(\rho,\rho)$ 
 commonly used in the  linear regression.
For  highly skewed data, $\lambda_j=\psi_j\sqrt{\gamma_j}$ is large, and  $\sigma_j^2=1/\gamma_j$ is close to $0$. We expect  that both
the residual sum of square error $\hat{S}_j$ from model \eqref{skewtdist} and $d_{\psi_j}$ are close to $0$. 
Under the gamma prior, $E_j=\text{diag}(  \frac{ 4 d_{\psi_j} }{\pi^2},0,\ldots,0  , 2\rho) \rightarrow E_j^* =\text{diag}(0,\ldots,0,  2\rho) $.   The marginal posterior distribution of $\gamma_j$
is approximately a gamma distribution with rate parameter $\rho+\hat{S}_j/2$ [this holds exactly if $E_j=E_j^*$, or if a flat prior is used for $\psi_j$].
The gamma prior can be quite informative when $\hat{S}_j$ is relatively small compared to $\rho$. 
We do not use the Jeffreys prior $\pi(\gamma_j)\propto \gamma_j^{-1}$ since the matrix $D_j+E_j$ can be nearly singular for highly skewed data, causing numerical problems.

\subsubsection{Prior and posterior distributions for $\nu_j$}\label{prepostnu}
We firstly derive the PC prior for $\nu_j$.
Let  $f(x)$ and $h(x)$ denote respectively the PDF of the t distribution $t(\mu, \frac{\nu-2}{\nu}\sigma^2,\nu)$ and normal distribution $N(\mu,\sigma^2)$. It is easy to show \cite{tang:2019} that
  $\int f(x)\log f(x)dx =\log\Gamma (\frac{\nu+1}{2})-\log\Gamma (\frac{\nu}{2}) -\frac{\nu+1}{2}[\Psi(\frac{\nu+1}{2})-\Psi(\frac{\nu}{2})]-\frac{1}{2}\log|\sigma^2|
  -\frac{1}{2}\log(\nu_j-2)-\frac{1}{2}\log(\pi)$,
and $\int f(x) \log h(x)dx = -\frac{1}{2}\log(2\pi)-\frac{1}{2}\log|\sigma^2|-\frac{1}{2}$.
 The Kullback-Leibler distance between the two distributions is
$$ KL (\nu)= \frac{1}{2}[1+\log(\frac{2}{\nu-2})] +\log\Gamma (\frac{\nu+1}{2})-\log\Gamma (\frac{\nu}{2})-\frac{\nu+1}{2}[\Psi(\frac{\nu+1}{2})-\Psi(\frac{\nu}{2})].$$
By the definition of the PC prior \cite{simpson:2017}, $d(\nu)=\sqrt{2 KL(\nu)}$, and 
the PC prior density is 
$$\pi(\nu)\propto \varrho \exp[-\varrho\, d(\nu)]\, |\frac{\partial d(\nu)}{\partial{\nu}}|,$$
where $\Gamma(\cdot)$, $\Psi(\cdot)$  and $\Psi'(\cdot)$ are the gamma, digamma and trigamma functions, $b(\nu)=\Psi(\frac{\nu+1}{2})-\Psi(\frac{\nu}{2})$, 
 $$d(\nu)=\sqrt{1+\log(\frac{2}{\nu-2}) +2\log\frac{\Gamma (\frac{\nu+1}{2})}{\Gamma (\frac{\nu}{2})}-(\nu+1)\,b(\nu)} \,\,\text{ and }\,\,
|\frac{\partial d(\nu)}{\partial{\nu}}|= \frac{\frac{1}{\nu-2} +\frac{\nu+1}{2} [\Psi'(\frac{\nu+1}{2})-\Psi'(\frac{\nu}{2})] }{4d(\nu)}.$$

The posterior distribution of $\nu_j$  is given by
\begin{equation}\label{postnu} 
 \pi(\nu_j|\gamma_j, \psi_j,\betav_j, Y_{d},Y_c,Y_{o})\propto 
 \pi(\nu_j) \prod_{i=1}^{n_j}  t (y_{ij}^*;\nu_j) T_{\nu+1} \left[\lambda_j \,y_{ij}^*\sqrt{\frac{\nu_j+1}{\nu_j+y_{ij}^{*2}}}\right]I(\nu_j>\nu_l), 
\end{equation}
where $\omega_{ij}^2=\gamma_j^{-1}+\psi_j^2$, $y_{ij}^*= (y_{ij} - \zv_{ij}'\betav_j)/\omega_{ij}$, and  $\lambda_j=\psi_j\sqrt{\gamma_j}$.

\subsubsection{Posterior distribution of $(\wc_{ij},d_{ij})$}
The posterior distribution of $(\wc_{ij},d_{ij})$  given $(\nu_j, \betav_j,\psi_j,\gamma_j, d_{\psi_j}, \rho_j, \yv_i)$ is
\begin{eqnarray}\label{postdwy0}
\begin{aligned}
\text{pos}(d_{ij},\wc_{ij}) & \propto d_{ij}^{\frac{\nu+2}{2}-1} \exp[-d_{ij}\frac{\nu+\wc_{ij}^2 }{2}] \exp[-\frac{\gamma_jd_{ij}(y_{ij}-\zv_{ij}'\betav_j-\psi_j \wc_{ij})^2}{2}] I(\wc_{ij}>0)\\
               & \propto \left\{d_{ij}^{\frac{\nu+1}{2}-1} \exp[-d_{ij}\frac{b_d}{2}]\right\}
\left\{d_{ij}^{\frac{1}{2}}\exp[ -\frac{ d_{ij} V_w (\wc_{ij} -\mu_w)^2 }{2}  ]\right\} I(\wc_{ij}>0)\\
& \propto \left\{\left[ 1 + \frac{\frac{(\wc_{ij}-\mu_w)^2}{b_d/(b_aV_w)}}{b_a}   \right]^{-\frac{b_a+1}{2}}I(\wc_{ij}>0)\right\} 
          \left\{\left [ \frac{b_d^*}{2}\right]^{\frac{b_a+1}{2}} d_{ij}^{\frac{b_a+1}{2}-1} \exp[-d_{ij}\frac{b_d^*}{2}]        \right\},
\end{aligned}
\end{eqnarray}
where $y_{ij}^{**}=y_{ij}-\zv_{ij}'\betav_j$, $V_{w}=\gamma_j\psi_j^2+1$, $\mu_{w}= \frac{\gamma_j\psi_j y_{ij}^{**}}{V_{w}}$,
$b_a= \nu+1$, $b_d= \nu +\frac{\gamma_jy_{ij}^{**2}}{V_{w}}$, 
and $b_d^*=b_d + (\wc_{ij}-\mu_w)^2 V_w$. In Equation \eqref{postdwy0},
the marginal distribution of $\wc_{ij}$ is a positive t distribution $t^+(\mu_{w}, \frac{b_d}{b_a V_{w}}, b_a)$, and the conditional distribution of $d_{ij}$ given $\wc_{ij}$ is
$\mathcal{G}(\frac{b_a+1}{2},  \frac{ b_d^*}{2})$. They can be generated as 
\begin{eqnarray}\label{postdwy1}
d_{ij}^*\sim \mathcal{G}(\frac{b_a}{2},\frac{b_d}{2}), \wc_{ij}|d_{ij}^* \sim N^+(\mu_{w}, \frac{1}{d_{ij}^* V_{w}}) ,\, d_{ij} |\wc_{ij}  \sim \mathcal{G}(\frac{b_a+1}{2},  \frac{ b_d^*}{2}).
\end{eqnarray}
Note that in Equation \eqref{postdwy0}, the marginal distribution of $d_{ij}$  is not gamma.

\subsubsection{Generation of random variables from   $f(g)\propto g^{c-1} \exp(-bg)\exp(-\frac{a}{g})$ with $c>0$, $b>0$ and $a\geq0$}\label{ghpx}
We use the  acceptance and rejection algorithm.
A candidate $g^*$ is drawn from $\mathcal{G}(c,d)$ with PDF $h(g)\propto g^{c-1} \exp(-dg)$, where $r=b-d\geq 0$. Thus
$f(g)/h(g)=\exp(-r g -a/g) \leq \exp[-2\sqrt{r a}]$. We accept $g^*$ with probability $\exp[-(\sqrt{r g}-\sqrt{a/g})^2]$.
We set $\sqrt{a/r}=\text{E}(g^*)=c/d$. That is $r = b \frac{e-1}{e+1}$ and $d=\frac{2b}{e+1}$, where $e= \sqrt{1+4ab/c^2}$.
The acceptance rate is typically higher than $0.9$ in our numerical examples.
When $a=0$, we have $e=1$, $b=d$ and the acceptance is $1$. 

\subsubsection{Posterior distribution of $g$ and $h$ in steps PX1 and PX2}
The posterior distribution of $g$ under the Haar prior $\pi(g)\propto g^{-1}$ with Jacobian $g^{n_j-1}$ is given by
\begin{eqnarray}\label{postg}
\begin{aligned}
\text{pos}(g) &\propto g^{n_j-1} g^{-1} f(\nu_j, g\,d_{1j},\ldots, g\,d_{n_jj},\wc_{1j},\ldots, \wc_{n_jj}, \betav_j,\psi_j,\frac{\gamma_j}{g}, d_{\psi_j}, \rho_j|Y_o,Y_d,Y_c) \\
&\propto
 g^{\frac{n_j(\nu_j+1)-(n_0+1)}{2}-1} \exp\left[-g\frac{\sum_{i=1}^{n_j} d_{ij}(\nu_j+\wc_{ij}^2)}{2}\right] \exp\left[-\frac{ \gamma_j (n_0\rho_j + \frac{2d_{\psi_j}\psi_j^2}{\pi^2})}{g}\right].
\end{aligned}
\end{eqnarray}
The posterior distribution of $h$ under the Haar prior $\pi(h)\propto h^{-1}$ with Jacobian $h^{n_j-1}$ is given by
\begin{eqnarray*}\label{posth0}
\begin{aligned}
\text{pos}(h) &\propto h^{n_j-1} h^{-1} f(\nu_j,d_{1j},\ldots, d_{n_jj},h\,\wc_{1j},\ldots, h\,\wc_{n_jj}, \betav_j,\frac{\psi_j}{h}, \gamma_j,d_{\psi_j}, \rho_j|Y_o,Y_d,Y_c) \\
&\propto h^{n_j-2} \exp\left[-h^2\frac{\sum_{i=1}^{n_j} d_{ij}\wc_{ij}^2}{2}\right] \exp\left[-\frac{2  d_{\psi_j} \gamma_j \psi_j^2}{h^2\pi^2}\right].
\end{aligned}
\end{eqnarray*}
Therefore, the posterior distribution of $H=h^2$ is 
\begin{equation}\label{posth}
\text{pos}(H)\propto \text{pos}(h) \frac{1}{\sqrt{H}} \propto H^{\frac{n_j-1}{2}-1} \exp\left(-H \frac{\sum_{i=1}^{n_j} d_{ij} \wc_{ij}^2}{2}\right) \exp\left(-\frac{2 d_{\psi_j}\gamma_j\psi_j^2}{\pi^2H}\right).
\end{equation}
We can draw $g$ and $H=h^2$ using the method described in Appendix \ref{ghpx}.

\subsection{Justification of the FCS-MNAR algorithm}\label{fcsmnarapp}
The MCMC algorithm in Section \ref{seqregressionsec} can be summarized as below 
\begin{itemize}
\item Draw $(\betav_j\text{'s},\phi_j\text{'s}, Y_d, Y_c)$ from their posterior distribution given $Y_o$ until the MDA algorithm converges
\item Impute $Y_w$ given $(\betav_j\text{'s},\phi_j\text{'s}, Y_d, Y_c, Y_o)$.
\end{itemize}
The following variant of the algorithm is valid, but  less efficient since an additional step is needed to draw $(\betav_j^*,\phi_j^*)$\text{'s} 
\begin{itemize}
\item Draw $(\betav_j\text{'s},\phi_j\text{'s}, Y_d, Y_c)$ from their posterior distribution given $Y_o$ until the MDA algorithm converges
\item Impute $Y_w$ given $(Y_d, Y_c, Y_o)$. This can be done by drawing $(\betav_j^*,\phi_j^*)$\text{'s} given $(Y_d, Y_c, Y_o)$, and sampling $Y_w$ from their posterior distribution given $(\betav_j^*\text{'s},\phi_j^*\text{'s}, Y_d, Y_c, Y_o)$.
\end{itemize}
The idea underlying the FCS-MNAR algorithm is similar to the above variant except that the intermittent missing outcomes in the first step are imputed by FCS.

\bibliographystyle{wileyj}
\bibliography{glmimp} 

\begin{thebibliography}{10}
\providecommand{\url}[1]{\texttt{#1}}
\providecommand{\urlprefix}{URL }
\expandafter\ifx\csname urlstyle\endcsname\relax
  \providecommand{\doi}[1]{doi:\discretionary{}{}{}#1}\else
  \providecommand{\doi}{doi:\discretionary{}{}{}\begingroup
  \urlstyle{rm}\Url}\fi

\bibitem{rubin:1996}
Rubin DB. Multiple imputation after 18+ years. \emph{Journal of the American
  Statistical Association}  1996; \textbf{91}:473--89.

\bibitem{rubin:1987}
Rubin DB. \emph{Multiple Imputation for Nonresponse in Surveys}. New York: John
  Wiley Sons, Inc, 1987.

\bibitem{little:1996}
Little R, Yau L. Intent-to-treat analysis for longitudinal studies with
  drop-outs. \emph{Biometrics}  1996; \textbf{52}:1324 -- 33.

\bibitem{faucett:2002}
Faucett CL, Schenker N, Taylor JMG. Survival analysis using auxiliary variables
  via multiple imputation, with application to {AIDS} clinical trials data.
  \emph{Biometrics}  2002; \textbf{58}:37--47.

\bibitem{collins:2001}
Collins LM, Schafer JL, Kam CM. A comparison of inclusive and restrictive
  strategies in modern missing data procedures. \emph{Psychological Methods}
  2001; \textbf{6}:330 -- 51.

\bibitem{schafer:2003}
Schafer JL. Multiple imputation in multivariate problems when the imputation
  and analysis models differ. \emph{Statistica Neerlandica}  2003;
  \textbf{57}:19--35.

\bibitem{schafer:1997}
Schafer JL. \emph{Analysis of Incomplete Multivariate Data}. Chapman Hall,
  London, 1997.

\bibitem{2016:tang}
Tang Y. An efficient monotone data augmentation algorithm for multiple
  imputation in a class of pattern mixture models. \emph{Journal of
  Biopharmaceutical Statistics}  2017; \textbf{27}:620 -- 38.

\bibitem{2015a:tang}
Tang Y. An efficient monotone data augmentation algorithm for {B}ayesian
  analysis of incomplete longitudinal data. \emph{Statistics \& Probability
  Letters}  2015; \textbf{104}:146 -- 52.

\bibitem{raghunathan:2001}
Raghunathan TE, Lepkowski JM, {van Hoewyk} J, Solenberger P. A multivariate
  technique for multiply imputing missing values using a sequence of regression
  models. \emph{Survey Methodology}  2001; \textbf{27}:85 -- 95.

\bibitem{Bernaards:2007}
Bernaards CA, Belin TR, Schafer JL. Robustness of a multivariate normal
  approximation for imputation of incomplete binary data. \emph{Statistics in
  Medicine}  2007; \textbf{26}:1368 -- 82.

\bibitem{lee:2010b}
Lee KJ, Carlin JB. Multiple imputation for missing data: Fully conditional
  specification versus multivariate normal imputation. \emph{American Journal
  of Epidemiology}  2010; \textbf{171}:624 -- 32.

\bibitem{donneau:2015a}
Donneau AF, Mauer M, Molenberghs G, Albert A. A simulation study comparing
  multiple imputation methods for incomplete longitudinal ordinal data.
  \emph{Communications in Statistics - Simulation and Computation}  2015;
  \textbf{44}:1311 -- 1338.

\bibitem{buuren:2006}
{van Buuren} S, Brand JPL, Groothuis-oudshoorn CGM, Rubin DB. Fully conditional
  specification in multivariate imputation. \emph{Journal of Statistical
  Computation and Simulation}  2006; \textbf{76}:1049 -- 64.

\bibitem{buuren:2007}
{van Buuren} S. Multiple imputation of discrete and continuous data by fully
  conditional specification. \emph{Statistical Methods in Medical Research}
  2007; \textbf{16}:219 -- 42.

\bibitem{white:2011}
White IR, Royston P, Wood AM. Multiple imputation using chained equations:
  Issues and guidance for practice. \emph{Statistics in Medicine}  2011;
  \textbf{30}:377 -- 99.

\bibitem{casella:1992}
Casella G, George EI. Explaining the {G}ibbs sampler. \emph{The American
  Statistician}  1992; \textbf{46}:167 -- 74.

\bibitem{liu:2014}
Liu J, Gelman A, Hill J, Su Y, Kropko J. On the stationary distribution of
  iterative imputations. \emph{Biometrika}  2014; \textbf{101}:155 -- 73.

\bibitem{chen:2015}
Chen SH, Edward Hl. Behavior of the {G}ibbs sampler when conditional
  distributions are potentially incompatible. \emph{Journal of statistical
  computation and simulation}  2015; \textbf{85}:3266 --75.

\bibitem{verbeke:2014}
Verbeke G, Fieuws S, Molenberghs G, Davidian M. The analysis of multivariate
  longitudinal data: A review. \emph{Statistical Methods in Medical Research}
  2014; \textbf{23}:42--59.

\bibitem{tang:2017e}
Tang Y. Algorithms for imputing partially observed recurrent events with
  applications to multiple imputation in pattern mixture models. \emph{Journal
  of Biopharmaceutical Statistics {\normalfont DOI:
  10.1080/10543406.2017.1333999}}  2017; .

\bibitem{tang:2018}
Tang Y. Controlled pattern imputation for sensitivity analysis of longitudinal
  binary and ordinal outcomes with nonignorable dropout. \emph{Statistics in
  Medicine}  2018; \textbf{37}:1467 -- 81.

\bibitem{lee:2016}
Lee MC, Mitra R. Multiply imputing missing values in data sets with mixed
  measurement scales using a sequence of generalised linear models.
  \emph{Computational Statistics and Data Analysis}  2016; \textbf{95}:24--38.

\bibitem{Goldstein:2009}
Goldstein H, Carpenter J, Kenward MG, Levin KA. Multilevel models with
  multivariate mixed response types. \emph{Statistical Modelling}  2009;
  \textbf{9}:173 -- 97.

\bibitem{chmp:2010}
{CHMP}. \emph{EMA Guideline on Missing data in Confirmatory Clinical Trials
  (EMA/CPMP/EWP/1776/99)}. London: CHAMP, 2010.

\bibitem{ich:2017}
\emph{ICH E9 (R1) addendum on estimands and sensitivity analysis in clinical
  trials to the guideline on statistical principles for clinical trials}.
  {http://www.ema.europa.eu/docs/en\_GB/document\_library/Scientific\_guideline/2017/08/WC500233916.pdf},
  2017.

\bibitem{NRC:2010}
{National Research Council}. \emph{The prevention and treatment of missing data
  in clinical trials}. The National Academies Press: Washington, DC, 2010.

\bibitem{2013:mallinckrodta}
Mallinckrodt C, Roger J, Chuang-Stein C, Molenberghs G, Lane PW, O'kelly M,
  {\it et al}. Missing data: Turning guidance into action. \emph{Statistics in
  Biopharmaceutical Research}  2013; \textbf{5}:369 -- 82.

\bibitem{2013:carpenter}
Carpenter JR, Roger JH, Kenward MG. Analysis of longitudinal trials with
  protocol deviation: a framework for relevant, accessible assumptions, and
  inference via multiple imputation. \emph{Journal of Biopharmaceutical
  Statistics}  2013; \textbf{23}:1352 -- 71.

\bibitem{2016:tangb}
Tang Y. An efficient multiple imputation algorithm for control-based and
  delta-adjusted pattern mixture models using {SAS}. \emph{Statistics in
  Biopharmaceutical Research}  2017; \textbf{9}:116 -- 25.

\bibitem{nelder:1972}
Nelder JA, Wedderburn RWM. Generalized linear models. \emph{Journal of the
  Royal Statistical Society, Series A}  1972; \textbf{135}:370 -- 84.

\bibitem{McCullagh:1989}
{P McCullagh and J A Nelder}. \emph{Generalized Linear Models, 2nd Edition}.
  Chapman \& Hall, London, 1989.

\bibitem{gamerman:1997}
Gamerman D. Efficient sampling from the posterior distribution in generalized
  linear mixed models. \emph{Statistics and Computing}  1997; \textbf{7}:57 --
  68.

\bibitem{albert:1993}
Albert JH, Chib S. Bayesian analysis of binary and polychotomous response data.
  \emph{Journal of the American Statistical Association}  1993; \textbf{88}:669
  -- 79.

\bibitem{liu:1999}
Liu JS, Wu YN. Parameter expansion for data augmentation. \emph{Journal of the
  American Statistical Association}  1999; \textbf{94}:1264 -- 74.

\bibitem{liu:2000}
Liu JS, Sabatti C. Generalised {G}ibbs sampler and multigrid {M}onte {C}arlo
  for {B}ayesian computation. \emph{Biometrika}  2000; \textbf{87}:353--69.

\bibitem{vandyk:2015}
{Van Dyk} DA, Jiao X. Metropolis-hastings within partially collapsed gibbs
  samplers. \emph{Journal of Computational and Graphical Statistics}  2015;
  \textbf{24}:301--27.

\bibitem{lee:2017}
Lee KJ, Carlin JB. Multiple imputation in the presence of non-normal data.
  \emph{Journal of Computational and Graphical Statistics}  2017;
  \textbf{36}:606 -- 617.

\bibitem{azzalini:2003}
Azzalini A, Capitanio A. Distributions generated by perturbation of symmetry
  with emphasis on a multivariate skew t distribution. \emph{Journal of the
  Royal Statistical Society, Series B}  2003; \textbf{65}:367 -- 389.

\bibitem{AZZALINI:1985}
Azzalini A. A class of distributions which includes the normal ones.
  \emph{Scandinavian Journal of Statistics}  1985; \textbf{12}:171 -- 78.

\bibitem{liseo:2006}
Liseo B, Loperfido N. A note on reference priors for the scalar skew-normal
  distribution. \emph{Journal of Statistical Planning and Inference}  2006;
  \textbf{136}:373 -- 89.

\bibitem{branco:2013}
Branco MD, ~ MGG, Liseo B. Objective {B}ayesian analysis of skew-t
  distributions. \emph{Scandinavian Journal of Statistics}  2013;
  \textbf{40}:63-- 85.

\bibitem{bayes:2007}
Bayes CL, Branco MD. Bayesian inference for the skewness parameter of the
  scalar skew-normal distribution. \emph{Brazilian Journal of Probability and
  Statistics}  2007; \textbf{21}:141 -- 63.

\bibitem{gelman:2006}
Gelman A. Prior distributions for variance parameters in hierarchical models.
  \emph{Bayesian Analysis}  2006; \textbf{1}:515 -- 534.

\bibitem{huang:2013}
Huang A, Wand MP. Simple marginally noninformative prior distributions for
  covariance matrices. \emph{Bayesian Analysis}  2013; \textbf{8}:439 -- 452.

\bibitem{fernandez:1999}
Fernandez C, Steel MFJ. Multivariate student t regression models: pitfalls and
  inference. \emph{Biometrika}  1999; \textbf{86}:153 -- 67.

\bibitem{fonseca:2008}
Fonseca TC, Ferreira MAR, Migon HS. Objective {B}ayesian analysis for the
  {S}tudent-t regression model. \emph{Biometrika}  2008; \textbf{95}:325 -- 33.

\bibitem{simpson:2017}
Simpson D, Rue H, Riebler A, Martins TG, Sorbye SH. Penalising model component
  complexity: A principled, practical approach to constructing priors.
  \emph{Statistical Science}  2017; \textbf{32}:1--28.

\bibitem{tang:2019}
Tang Y. Monotone data augmentation algorithm for longitudinal data analysis via
  multivariate skew-t and skew-normal distributions. \emph{submitted}  2019; .

\bibitem{gilks:1992}
Gilks WR, Wild P. Adaptive rejection sampling for {G}ibbs sampling.
  \emph{Applied Statistics}  1992; \textbf{41}:337 -- 48.

\bibitem{damlen:1999}
Damlen P, Wakefield J, Walker S. Gibbs sampling for {B}ayesian non-conjugate
  and hierarchical models by using auxiliary variables. \emph{Journal of the
  Royal Statistical Society B}  1999; \textbf{61}:331 -- 44.

\bibitem{permutt:2016}
Permutt T. Sensitivity analysis for missing data in regulatory submission.
  \emph{Statistics in Medicine}  2016; \textbf{35}:876 -- 9.

\bibitem{buuren:2011}
{van Buuren} S, Groothuis-Oudshoon K. Mice: Multivariate imputation by chained
  equations in {R}. \emph{Journal of Statistical Software}  2011;
  \textbf{45}:1--67.

\bibitem{2014d:lu}
Lu K. Number of imputations needed to stabilize estimated treatment difference
  in longitudinal data analysis. \emph{Statistical Methods in Medical Research}
   2017; \textbf{26}:674 -- 90.

\bibitem{2016:tangc}
Tang Y. On the multiple imputation variance estimator for control-based and
  delta-adjusted pattern mixture models. \emph{Biometrics}  2017;
  \textbf{73}:1379 -- 87.

\bibitem{EMA:2011}
{European Medicines Agency}. \emph{Guideline on clinical investigation of
  medicinal products in the treatment of depression}. 2011.

\bibitem{seaman:2016}
Seaman SR, Hughes RA. Relative efficiency of joint-model and
  full-conditional-specification multiple imputation when conditional models
  are compatible: The general location model. \emph{Statistical Methods in
  Medical Research \normalfont{DOI: 10.1177/0962280216665872}}  2016; .

\bibitem{hippel:2013}
{von Hippel} PT. Should a normal imputation model be modified to impute skewed
  variables? \emph{Sociological Methods and Research}  2013; \textbf{42}:105--
  38.

\bibitem{schenker:1996}
Schenker N, Taylor JMG. Partially parametric techniques formultiple imputation.
  \emph{Computational Statistics \& Data Analysis}  1996; \textbf{22}:425 --
  46.

\end{thebibliography}
\end{document}